\documentclass[11pt,reqno]{article}

\usepackage{amssymb,amsmath,amsthm}
\usepackage{latexsym}
\usepackage{graphicx}
\usepackage{ctable,multirow}
\usepackage[margin=2.5cm]{geometry}

\usepackage{epsfig}
\usepackage{lscape}
\usepackage{ctable,multirow}
\usepackage{colortbl,color,array} 

\newcommand{ \U }{\overline{U}}

\newcommand{ \bmu }{\mbox{\boldmath ${\mu}$}}
\newcommand{ \beps}{\mbox{\boldmath ${\epsilon}$}}

\newcommand{ \by }{\mbox{\boldmath ${y}$}}
\newcommand{ \bX }{\mbox{\boldmath ${X}$}}
\newcommand{ \bY }{\mbox{\boldmath ${Y}$}}
\newcommand{ \bz }{\mbox{\boldmath ${z}$}}
\newcommand{ \bZ }{\mbox{\boldmath ${Z}$}}

\newcommand{ \X }{ \mbox{\boldmath ${X}$}}
\newcommand{ \Z }{ Z }
\newcommand{ \Vb }{\mbox{\boldmath ${\beta}$}}
\newcommand{ \wVb }{\widehat{\mbox{\boldmath ${\beta}$}}}
\newtheorem{theorem}{Theorem}

\begin{document}

\baselineskip=0.65cm

\title{A Consistent Direct Method for Estimating Parameters \\in Ordinary Differential Equations Models}
\author{Sarah E. Holte  \\
Division of Public Health Sciences\\
Fred Hutchinson Cancer Research Center \\
1100 Fairview Ave North, M2-B500 \\
Seattle, WA 98109, USA  \\
email: \texttt{sholte@fhcrc.org}
}

\maketitle


\begin{abstract}
Ordinary differential equations provide an attractive framework for modeling temporal
dynamics in a variety of scientific settings. We show how consistent estimation for
parameters in ODE models can be obtained by modifying a direct (non-iterative) least
squares method similar to the direct methods originally developed by Himmelbau, Jones
and Bischoff. Our method is called the bias-corrected least squares (BCLS) method since
it is a modification of least squares methods known to be biased. Consistency of the
BCLS method is established and simulations are used to compare the BCLS method to other
methods for parameter estimation in ODE models.
\end{abstract}


\section{Introduction}
Ordinary differential equations (ODEs) have been used extensively in a wide variety of
scientific areas to describe time varying phenomena. Physicists have used equations of
this type to understand the motion of objects for centuries. ODE's are used in nearly
every field of engineering and chemists use ODE models to evaluate chemical reactions.
Ecologists have used ODE's to understand the dynamics of animal and plant populations
\cite{Fre:80}, meteorologists have used them to understand and predict patterns in the
weather and atmosphere \cite{Lor:63}, and economists have used them to understand and
predict patterns in financial markets \cite{Zha:05}.  More recently, biologists and
health scientists have begun to use differential equations to make sense of observed
nonlinear behaviors in their fields of study. One area where the use of differential
equations is well integrated with data analysis is the evaluation of drug metabolism, the
study of pharmacokinetics and pharmacodynamics (PK/PD).  Studies of how drugs are
metabolized are part of the development of nearly all compounds available for use in
humans today. Reviews of the use of differential equations in both individual and
population level PK/PD include \cite{Csa:06,Dan:08,Dar:07,Mag:03,She:00}. Software
packages for parameter estimation for population PK/PD models are available in both SAS
\cite{Gal:04} and Splus \cite{Tor:04} . In other areas of biology and clinical research,
ordinary differential equation (ODE) models have been used as a tool in exploring the
etiology of HIV and Hepatitis B and C infections and the effects of therapy on these
diseases. In \cite{Ho:95,Wei:95,Per:96,Per:97} ordinary differential equations were used
to analyze the temporal dynamics of HIV viral load measurements in AIDS patients. The
results had a tremendous scientific impact, revealing that the HIV virus replicates
rapidly and continuously, in spite of the prolonged interval between infection and
development of AIDS.

In some cases, ODE's are used for theoretical purposes only; to provide ``what if"
scenarios to be further studied by experimental scientists. However, more and more, ODE's
are used in combination with observed data for parameter estimation and statistical
inference of parameters in ODE models. In most cases, the parameters of interest appear
in a nonlinear form in the likelihood or cost equation.  Estimates are often obtained
using standard nonlinear least squares (NLS) estimation techniques.  One drawback of the
NLS estimation technique using conventional gradient based optimization methods such as
Gauss-Newton or Levenberg-Marquart is the instability of the procedure since estimates
are obtained by an iterative numerical procedure (\cite{Pre:86}, Chapter 10). NLS
procedures may fail to converge to global minima depending on the choice of starting
values for iteration, particularly if the least squares solution has multiple local
minima or saddle points. Global optimization methods are often available, however many
analysts continue to rely on algorithms such as nonlinear least squares that require a
choice of starting values. Certain likelihood surfaces or cost functions may be
exceptionally complex with ridges which tend to appear near bifurcation values of the
parameters in the ODE model.  Furthermore, it is well known that nonlinear regression
estimates may be biased \cite{Cook:86} when small samples are used for estimation.

When parameters in an ODE model appear linearly in the vector field which defines the ODE
system, it is possible to estimate the parameters either iteratively (e.g., nonlinear
least squares) or non-iteratively. In this paper we will use the term {\em direct} to
refer to non-iterative methods. Direct methods can be based on differentiation or
integration and are generally combined with a least squares estimation step. Direct
methods based on integration were originally described in \cite{HimJon:67} and are
further developed in \cite{Bard:74,Foss:71,Hosten:79,Jac:74}. In general, the ODEs are
transformed into integral equations and the integrals are approximated via methods of
quadrature which yields algebraic equations that can be solved for the approximate
parameters using a linear least squares approach. Differentiation methods are described
in \cite{Bard:74,Hosten:79,SwaBre:75,Varah:82} and involve constructing a cubic spline
using the observed data for each dependent variable. Values for the unknown parameters
are estimated using linear least squares, i.e., minimizing the differences between the
derivatives of the spline functions and corresponding model equations. Both the
differentiation and integration approaches result in linear least squares problems which
can be solved without iteration. When there is negligible measurement error and equally
spaced values of the dependent variable, the two approaches provide similar estimates
\cite{Wikstrom:97a}. In the case where these two assumptions are not met, the integration
approach is preferred over the differentiation approach \cite{Wikstrom:97b}.
Differentiation requires differencing of observed data, an approach which is known to
inflate errors in the data.  In contrast, integration involves sums of observed data; a
smoothing approach which can reduce errors. An approach which relies on integration
allows estimation of the initial size of the states of the system if these are not known.

There has been very little investigation of the statistical properties of parameter estimates in ODE models
obtained by direct methods, possibly since it is known that parameter estimates obtained via these direct
methods do not have optimal statistical properties \cite{Hosten:79}.
Specifically, the estimates are not unbiased, efficient, or
consistent.  While these early direct methods did not have optimal statistical properties it
is possible that adjustments can be identified so that the resulting direct method has
desirable statistical features.

Recently, Liang and Wu \cite{Lia:08} describe a
direct method based on differentiation, local smoothing of the data
and linear least squares regression of correlated quantities, which they
refer to as pseudo-least squares (PsLS) method.  They show that under
reasonably general conditions their PsLS estimate of parameters
in an ODE model is consistent with an asymptotically normal distribution.
However, choice of bandwidth for the kernel smoothing required by this
method is critical and may require further
research in order to provide the most accurate and efficient results.
Ramsey and colleagues \cite{RamSil:05,RHCC:07}
have used functional estimation or principle differential analysis (PDA)
approaches to replace state variables with their (smoothed) estimates, and proceed
with estimation of parameters using estimated values of the solutions
to the system of ODEs. These methods, like those of Liang and Wu, use differentiation rather than integration.

In this paper we introduce a direct algorithm similar to those developed by
or based on the work of Himmelbau, Jones and Bischoff \cite{HimJon:67}.
Our approach incorporates a bias correction factor so that the method is consistent, and
does not require smoothing or functional estimation. As such it retains the simplicity
of early direct methods \cite{Bard:74,Jac:74,Foss:71,HimJon:67,Hosten:79,SwaBre:75,Varah:82}
but has many of the desirable statistical properties of the more recently developed direct
methods \cite{Lia:08,RHCC:07}. Our method relies
on integration similar to the methods described in \cite{Bard:74,Foss:71,HimJon:67,Hosten:79,Jac:74}.
We refer to the proposed method as the {\it bias-corrected least squares} method (BCLS)
and the focus of this paper is
to evaluate the statistical properties of the estimates produced by the BCLS method.
Since the BCLS method relies on integration
rather than differentiation, estimation of the initial states of the system in addition
to parameters appearing linearly in the system is possible.
The method is applied to two examples and its properties contrasted with those of the NLS
and PsLS methods with simulation studies. A proof of the consistency of the method is
provided in the appendix.

\section{The Bias-Corrected Least Squares Method}

\subsection{ Mathematical and Statistical  Model Specification }

The bias-corrected least squares method applies to data, ${\bf y}({\bf t})$, on observations with time varying
expectation given by the solutions of system of  $q=1,\cdots, s$ ODEs, observed
at ${\bf t}=(t_0,\cdots,t_n)$, $n$ observation times. It is not necessary that all $s$ compartments be
sampled at the same times, but for ease of notation it is assumed that this is the case in this work.
The statistical and mathematical model assumptions for the the data
${\bf y}({\bf t})$ with distribution ${\bf Y(t)}$ are summarized as follows:
\begin{itemize}

\item[(A.1)]  : Specification of the mean structure via a mathematical model
   \begin{itemize}
   \item There exists a vector of functions
  ${\bf X}(t)=\{X_1(t),\cdots,X_s(t)\}$ which satisfies the
  system of differential equations:
  \begin{eqnarray}
  \frac{dX_q}{dt}=\sum_{k=1}^{m_q} \beta_{q,k} h_{q,k}(X_1,\cdots,X_s), \ \
  X_q(t_0)=X_{q,0}, \ \ q=1,\cdots,s \ \
  \label{geneq}
  \end{eqnarray}
  such that that the expectation, $E\{{\bf Y}(t)\}$ is equal to $ {\bf X}(t)$.
    \end{itemize}

\vskip-.1in
\item[(A.2)] : Specification of variance/covariance structure
 \begin{itemize}
     \item $Y_q ({\bf t}) = X_q ({\bf t}) + \beps_q,\ \ \beps_q \sim({\bf 0},\Sigma_q), \ \ q = 1,\cdots,s.$
     \item $\Sigma_q$ is a diagonal matrix with constant diagonal entries $\sigma_q$, \ \ $q = 1,\cdots,s$.
     \item $ \beps_r$ and $\beps_q$ are independent for all $1 \leq r < q \leq s$.
   \item  $\mbox{var}(Y_i)$ and $\mbox{var}\{h_{q,k}(Y_i)\}$ are bounded by a common bound $B$ for
all $i,q$, and $k$.
   \end{itemize}

\vskip-.1in
  \item[(A.3)] : Data sampling requirements
  \begin{itemize}
  \item The maximum interval length defined by the sampling times
is $O(n^{-1})$.
  \item At least one of the compartments is not in
equilibrium throughout the entire time course of data collection .
  \end{itemize}
\end{itemize}

The values $\beta_{q,k}$ and in some cases the initial conditions $X_{q,0}$ are
parameters that will be estimated from the data. The functions, $h_{q,k}$, are any
(linear or nonlinear) continuous functions of the states of the system, which together
with the parameters, $\beta_{q,k}$, and initial conditions, $X_{q,0}$, define the vector
field of the differential equations relating the states $\bX$ to their time derivatives.
For the $q^{th}$ state, the value $m_q$ is the number of terms in the expression which
specifies $\frac{d\bX_q}{dt}$. Note that the vector field defining the system of
differential equations can be nonlinear in the states, i.e., $h_{q,k}(\bX)$ is not
necessarily linear in the states, but must be linear in the parameters in order for the
bias-corrected least squares method (or any direct least squares based method) to be
applicable.  Thus the BCLS method applies to a wide variety of nonlinear systems of
ODE's.

Condition (A.1) provides the relationship between the time-varying expectation of the data and a system of ODEs.
Condition (A.2) requires that conditional on the expected value, observations from different
compartments of the system of ODE's are independent.  This requirement is not overly restrictive, since
the use of ODE's is intended to capture the relationships between the observed compartments.
Condition (A.2) also specifies that the variance of observations from different compartments can differ.
The second part of condition (A.3) is included to prevent non-identifiability, but
does not insure identifiability of the estimation procedure.

The key component of the BCLS method is the use of bias correction functions which are needed to account for bias which is introduced when
making nonlinear transformations of random variables. These bias correction functions are used as weights in the least squares estimation.
These functions can be defined in a number of ways.
For a function $h = h_{q,k}$,
identify the difference, $E[h\{\bY(t)\}] - h\{E[\bY(t)]\}$ and then subtract it from $h$ to define $h^{\star}$ as follows:
\begin{equation}\label{b1}
h^{\star}\{ \bY(t)\} =h \{ \bY(t) \} - \left(  E[h\{\bY(t)\}] - h\{E[\bY(t)]\}    \right).
\end{equation}
An alternative is to define $h^{\star}$ as:
\begin{equation}\label{b2}
h^{\star}\{ \bY(t)\} =  h \{ \bY(t) \}  \frac { h\{E[\bY(t)]\}   }  { E[h\{\bY(t)\}] }
\end{equation}
Taking expectations of both sides of  (\ref{b1}) or (\ref{b2}) shows that either definition satisfies
\begin{eqnarray}\label{ehstar}
E[h_{q,k}^{\star}\{{\bf Y}(t)\}]=h_{q,k}\{ \bX(t) \}, \ \ q=1,\cdots,s,\  k=1,\cdots,m_q . \ \
\end{eqnarray}
Any function, $h^{\star}$ which satisfies (\ref{ehstar}) can be used to weight the linear regression step
of the estimation procedure to eliminate bias in the resulting point estimates.  The previous two examples
show that such a function always exists; the most efficient form of $h^{\star}$ for a specific model generally depends on
the error structure of the data (e.g additive vs multiplicative).

The following notation will be used:
Let $y_{q,i}$, $q=1,\cdots,s$, $i=1,\cdots,n$, denote an observation with expectation given
by $X_q$ in Eq. (\ref{geneq}) at time $t_i$;
$y_q({\bf t})$ denote the vector of all observations from the $q^{th}$
compartment, $X_q({\bf t})$; ${\bf y}(t_i)$ denote the vector of observations on all
compartments at a single observation time, $t_i$; and ${\bf y}({\bf
t})$ denote the matrix of observations on all compartments at all
observation times. In general, lower case letters represent observed
data, and upper case represent the associated random variables or state variables of the system of ODE's.
In the following sections, when direct estimation of parameters in an ODE model is performed without the
bias-correction  we will refer to the resulting point estimates as simply least squares (LS) estimates or
least squares without bias correction.

\subsection{ Estimation Algorithm }
The BCLS method is designed to simplify a
nonlinear regression problem by reducing it to a linear regression problem.
This is achieved by transforming the system of differential equations (\ref{geneq})
into a system of integral equations and treating the transformed system as a
statistical model
for a collection of linear regressions with covariates which approximate the integrals.
Approximations to these covariates are obtained using
the observed data.
Specifically, Eq. (\ref{geneq})
is equivalent to the system of integral equations:
\begin{eqnarray*}
  X_q(t) = \sum_{k=1}^{m_q} \beta_{q,k}
    \int_{t_0}^{t} h_{q,k}\{ \bX(\tau)\} d\tau+ \X_{q,0}\ \
    q=1,\cdots, s.
\end{eqnarray*}
so that (A.1) implies
\begin{eqnarray}
  E\{Y_q(t)\} & = &  \sum_{k=1}^{m_q} \beta_{q,k}
    \int_{t_0}^{t} h_{q,k}\{ \bX(\tau) \} d\tau+ \X_{q,0}\ \
    q=1,\cdots, s.
\label{ey}
\end{eqnarray}
The expectations in Eq. (\ref{ey}) suggest
$s$ linear regression models with response variables (outcomes) $y_q({\bf t})$ and
covariates which approximate the integrals $\int_{t_0}^{t}  h_{q,k}\{ \bX(\tau) \} d\tau $ ,
$k=1,\cdots,m_q$, $q=1,\cdots, s$. If an intercept is included in the regression model it provides an estimate
for the initial condition, $X_q(t_0)$, and the estimated coefficients of the covariates
provide estimates of the parameters $\beta_{q,1}, \cdots, \beta_{q,m_q}$.
If the initial value $ y_q(t_0)$ is known, the response variable
can be defined as $y_q({\bf t}) - y_q(t_0)$ and linear regression without
intercept can be used.
For ease of notation, we now drop the subscript $q$ in the response
variable and describe each of the $s$ states separately.

To obtain covariates for the regression analysis, any method of integral approximation can be used.
For the examples and simulations in this manuscript, the trapezoid method is used
to approximate the integrals
$ \int_0^{t_i}  h_{k}\{ \bX(\tau) \}  d\tau $ and construct
covariates, $z_k(t_i)$, defined by:  $z_k(t_0)=0$ and for $k=1,\cdots,m$ and $i=0,\cdots,n$,
\[  z_k(t_i) =  {\displaystyle \sum_{j=1}^i }
 \frac{ [  h_k^{\star} \{ {\bf y}(t_{j-1}) \}  + h_k ^{\star} \{ {\bf  y}(t_j)
    \} ] *  (t_{j}-t_{j-1}) }{2}
\]
where $h_k^{\star}$ is the bias correction adjustment which satisfies (\ref{ehstar}), e.g (\ref{b1}) or (\ref{b2}).
Let $z_k({\bf t})$ denote the vector, $\{z_k(t_0),\cdots,z_k(t_n)\}$, and ${\bf z(t)}$ denote the matrix
with columns $z_k({\bf t})$, $k=1,\cdots,m$.
Although the trapezoid rule for integral approximations is used for the examples and simulations presented in this work,
many methods are available for integral approximation (\cite{Pre:86}, Chapter 4).
The most simple integral approximation is defined by
using only the left endpoints of the intervals and can be described using the
observed data as follows:
\[  z_k(t_i)=  {\displaystyle \sum_{j<i} }
  h_k^{\star}\{ {\bf y}(t_{j-1}) \}  * (t_{j}-t_{j-1})
\]
Note that if only the left hand endpoint is used in the definition of $z_k({\bf t})$,
the covariates are independent of the corresponding response, since the
definition of $z_k(t_i)$ depends only on data observed at time points $t_0,\cdots,t_{i-1}$
and the $i^{th}$ response is the data observed at time $t_i$.
Although we use this simple approximation to demonstrate consistency (Appendix 1),
in practice the trapezoid rule produces more
accurate estimates.

We can now define the linear regression model as
\[ Y({\bf t}) = {\bf Z(t)}\ \Vb + \beps, \ \ \  \beps \sim \ (0,\Sigma^2) \]
where $\Sigma^2$ is the variance/covariance matrix for the residual errors.
If only the left hand endpoint of each interval is used in the definition of $z_k({\bf t})$,
Condition (A.2) guarantees that $\Sigma$ is a diagonal matrix, although the
variances associated with each state are not assumed to be the same for all states.
In this work, we assume that $\Sigma$ is known or will be obtained using an
independent method.  These variances are usually needed to construct the
bias correction functions, $h_k^{\star}$.

To estimate $\Vb$, linear regression with response $y({\bf t})$
( or $y({\bf t}) - y(t_0)$ ) and the covariate matrix ${\bf z(t)}$ is performed
so that
  \begin{eqnarray}
  \wVb = ({\bf z(t)}'{\bf z(t)})^{-1}{\bf z(t)}'y({\bf t}) \ \
  \label{linreg}
  \end{eqnarray}
In \cite{Lia:08}, the authors refer to regression analysis of this type as pseudo-least squares,
(PsLS) since the minimization that occurs with the linear regression algorithm is not minimizing
a true likelihood. The same is true of bias-corrected least squares.
\vskip0.2in

To demonstrate consistency of the BCLS method we rely on the following Theorem.
{\small
\begin{theorem}
Assume conditions (A.1)-(A.3) (from Section 2.1 ) are
satisfied and that $\wVb_{n}$ is the unique root of the estimating function,
defined by \ \ $\displaystyle \U_{n}(\Vb)=\frac{1}{n}
\sum_{i=1}^n \bz_i^T \left\{ \by_i - \bz_i  \Vb \right\}=0$ .\ \
Then:
\begin{itemize}
\item[I]  $\U_{n}(\wVb_n)$ is a consistent estimator of zero.
\item[II] $E\{\frac{1}{n} \sum_{i=1}^n \bZ_i^T \bZ_i \}$ is bounded and
non-zero.
\item[III]  $ \mbox{var}\{\frac{1}{n} \sum_{i=1}^n \bZ_i^T \bZ_i\} \rightarrow 0$  as $n \rightarrow \infty$.
\end{itemize}
Thus, the linear least squares estimate, $\wVb = ({\bf z(t)}'{\bf z(t)})^{-1}{\bf z(t)}'y({\bf t})$
is a consistent estimator of $\Vb$.
\end{theorem}
}

The final conclusion, that the three conditions listed in Theorem 1
imply consistency of estimator $\wVb$ from standard linear regression
follows from results in \cite{Cro:86}.
The key technical issue is that the derived covariates, ${\bf z}({\bf t})$, are
functions of the observed data rather than of the state means and are therefore
measured with error, and that the covariates themselves are correlated with each other.
However, contrary to the classical measurement error
setting the variance of the covariates is $O(n^{-1})$ and therefore
bias due to error in the covariates is asymptotically eliminated.  It will be shown in the appendix that the BCLS method
satisfies condition [I-III].

\section{Example: A single compartment nonlinear ODE }

\subsection{Data Analysis: Growth colonies of paramecium aurelium}
To illustrate the method using a model defined by a nonlinear
differential equation with a single compartment, data on the growth of four colonies of the
unicellular organism, paramecium aurelium, were analyzed.  These data are described in \cite{Dig:90,Gau:34}.
In this example a differential equation represents the size of the population of
the colony of paramecium over time.  We
assume that the data on colony size, $y({\bf t})$, follows a log-normal distribution, with
\[
\log\{ Y({\bf t}) \} \  = \  \log\{ X({\bf t})\} + \beps,\ \beps \   \sim \  i.i.d.\;  N(0,\sigma^2)
\]
where $X(t)$ satisfies
the standard logistic growth curve described by the nonlinear differential
equation
\begin{eqnarray}
  \frac{dX}{dt}=X (a- b X), \ \  \X(0)=y_0
  \label{pareq}
\end{eqnarray}
The parameter  $a$ is the growth
rate per capita, $\frac{a}{b}$ is the carrying capacity of the
population, and $y_0$ is the initial size of the populations which
is known by design; $y_0=2$ in all four data sets.
A log transform is applied to the data
for analysis and parameter estimation so that a log-transform of the differential equation
(\ref{pareq}) is required.
\begin{eqnarray}
  \frac{d\{\log(X) \} }{dt}=a - b X, \ \  \log\{X(0)\}=\log(y_0).
  \label{logpareq}
\end{eqnarray}

To perform the BCLS method on the growth colony data,
equation (\ref{logpareq}) is rewritten as the integral equation
\begin{eqnarray*}
  \log\{ X(t) \} -\log(y_0) = a \int_0^{t} d\tau- b \int_0^{t} h\{ X(\tau)\} d\tau.
\end{eqnarray*}
where $h\{X\}= X$. The first covariate $\int_0^{t} d\tau$ is simply
${\bf t}$ and no weighting is necessary for this integral approximation.
To determine the correct form of $h^{\star}$ for computing the integral
approximation for  $\int_0^{t} h\{X(\tau)\} d\tau = \int_0^{t} X(\tau) d\tau$ we calculate
$E[h\left\{Y(t_i)\right\}] = E[Y(t_i)] = X(t_i) e^{\frac{\sigma^2}{2}}$
since the data, $y({\bf t})$ follows a log-normal distribution. If $h^{\star}(X) = X e^{-\frac{\sigma^2}{2}}$ it follows
that $E[h^{\star}\{Y(t_i)\}] = E[ Y(t_i)\  e^{-\frac{\sigma^2}{2}}] = X(t_i)\  e^{\frac{\sigma^2}{2}}\ e^{-\frac{\sigma^2}{2}} = X(t_i)$
as required by (\ref{ehstar}). Therefore, to approximate the
integral, $ \int_0^{t} X(\tau) d\tau $, we
define the covariates $z({\bf t})=\{z(t_0),z(t_1),\cdots,z(t_n)\}$ using the trapezoid rule for integral approximation:
$z(t_0)=0$ and for $i=1,\cdots,n$
\begin{eqnarray*}
 z(t_i) & = & e^{\frac{-\sigma^2}{2}} \left\{ \sum_{j=1}^{i-1} \frac{(y_j+y_{j-1})*(t_j-t_{j-1})}{2} \right\}
\end{eqnarray*}
Finally, since  the initial value, $y_0$, is known, the response
for the regression model for the BCLS method is given by $y'({\bf t})=\log\{y({\bf t})\}-\log\{y_0\}$.

To obtain the BCLS estimates of $a$ and $b$ we
use standard linear regression and the following model without intercept.
\begin{eqnarray}
y'({\bf t}) \sim a\ {\bf t}+ b\ z({\bf t}) + \beps,
\quad  \beps \sim i.i.d. \;  N(0,\sigma^2 ).
\label{regmod}
\end{eqnarray}
The estimated coefficient of ${\bf t}$ will provide an unbiased estimate of $a$ and the
estimated coefficient of $z({\bf t})$ will provide an unbiased estimate of $b$.  If the
weights, $e^{\frac{-\sigma^2}{2}}$, are not used in the construction of the covariate, $z({\bf t})$, the
resulting estimates of one or both of the parameters $a$ and $b$ would likely be biased.
If the initial value of the system, $y_0$, is unknown, it can be treated
as an additional parameter and
can be estimated by including an intercept in the regression model (\ref{regmod}) and
using $log\{y({\bf t})\}$ as the response.

An estimate of the measurement error, $\sigma$, was obtained by fitting a natural spline
with 3 degrees of freedom to the data (the ns() macro in R version 2.9.2). The residuals
from that fit were used to obtain an estimate of $\sigma$ of approximately $0.23$ which
was used to construct the regression covariates $z({\bf t})$. Next, the parameters $a$
and $b$ were estimated for the four different data sets on colonies of paramecium
aureilium using the LS without bias correction and the BCLS method. For nonlinear least
squares regression, estimates of the parameters from the BCLS method were used as
starting values for the Marquart estimation algorithm. Parametric bootstrap techniques
were used to obtain 95\% confidence intervals \cite{Efr:93}.

The estimated values of $a$ and $b$ and their 95\%  parametric bootstrap confidence intervals
are shown in Table 1.  Non-parametric bootstrap
confidence intervals were also obtained for the BCLS method, since non-parametric
bootstrap does not require an analytic or
numerical solution to equation (\ref{pareq}). The data sets and fitted
trajectories from both the NLS and BCLS estimation procedures
are shown in Figure 1.  The estimates and confidence
intervals obtained from the two methods are quite similar, and Figure
1 suggests that both methods produce estimates of the underlying
trajectories  which fit the data well.  For these four data sets, the BCLS method
produced virtually the same results as the LS method without bias correction (results not shown).

\subsection{Simulations}
To compare the BCLS method to NLS and the PsLS method described in \cite{Lia:08} for
accuracy and efficiency we simulated data from the statistical model
\begin{eqnarray*}
\log\{Y ({\bf t})\} & = &  \log\{X({\bf t})\} + \beps,
\quad \beps  \sim i.i.d \;  N(0,\sigma^2)
\end{eqnarray*}
where  $X({\bf t})$ is the solution of equation (\ref{pareq}) with
parameters set as follows: $a=0.8$, $b=0.0015$, and $y_0=2$.  These
parameters were chosen based on the fits to the four data sets shown
in Table 1. A range of values for residual error, $\sigma$, were evaluated in
order to demonstrate the increase in bias in
the estimate of the parameter $b$ using LS without
adjusting the covariate $z(t)$ as described in the previous section.
Two sets of observations times were
evaluated; Each consisted of observation times between $t=0$ and
$t=20$.  The first set of observation times consisted of 21 total
observations, evenly spaced 1 unit apart between 0 and 20.  The
second set of observation times consisted of 201 total observations,
evenly spaced 0.1 units apart between 0 and 20.
Individual simulations we conducted for
each of the two sets of observation times and each of the following values
of $\sigma$, $(0.2,0.4,0.6,0.8)$ for a total of 8 simulation studies.

For each combination of observation
times and values of $\sigma$, the data were
simulated 1000 times and the BCLS, LS (without weights to adjust for bias), PsLS and NLS
estimates of $a$ and $b$ were used to estimate model parameters.  To evaluate the PsLS method the local polynomial smoothing macro locpoly()
from the R KernSmooth package was used to obtain
the smoothed state and derivative curves.  A bandwidth of
order $\log(nt)^{-1/10}*nt^{-2/7}$ was chosen for both state and
derviative estimation as suggested in \cite{Lia:08}, where $nt$ is the sample size in each simulation.  A number of other choices for bandwidth we considered,
including the standard choice of order $nt^{-1/5}$.  The choice of bandwidth used in the simulations was made based on
selecting the bandwidth that produced the least biased estimates.

Figure 2 depicts the estimated values of the parameters $a$ and $b$ using the LS method (dotted lines), the BCLS method (combined dash and dotted lines), the NLS method (solid line) and the PsLS method (dashed line).  No bias correction is necessary for the integral approximation of the covariate used to estimate the parameter $a$ so that bias-corrected and unadjusted least squares estimates of $a$ are identical.  Panels A and B of Figure 2 show how bias in the
estimated parameters varies with measurement error in the simulated data when 21 equally
spaced time points between $t=0$ and $t=20$ are used; Panels C and D depict the same when the sample size is increased to 201 time points.
The Monte Carlo means and standard errors for the estimated parameters from each of the simulations with 21 equally spaced observations  are shown in Table 2.

As expected, Figure 2, panels B and D show that bias increases in the LS estimate of the
parameter $b$ as simulated measurement error, $\sigma$, increases from 0.2 to 0.8 when
bias correction is not used in covariate definition. Also as expected, the bias in the LS
estimate of $b$ without bias-correction is essentially unaffected by sample size (Figure
2, Panels C and D). With a total of 21 time points, the BCLS method appears to
underestimate the parameter $a$ slightly, likely due to errors in integral approximation
of a convex function with relatively sparse measurements.  The NLS method appears to
overestimate that same parameter, probably due to the bias associated with nonlinear
regression \cite{Cook:86}.  The bias in the BCLS and NLS estimates decreases
substantially when sample size is increased to 201 timepoints(Panes C and D). Figure 2
also shows that bias in PsLS estimates of both $a$ and $b$ changes as $\sigma$ increases,
indicating the constant in the choice of bandwidth affect bias in parameters estimated
using this method. . An increase in sample size from 21 time points to 201 time points
reduces the bias in the estimate of the parameter $a$ from the strongly consistent PsLS
method \cite{Lia:08}, however, bias in the estimate of parameter $b$ seems to shift but
not decrease, possibly due to an improper choice of bandwidth. Additional simulations
with larger samples sizes showed this bias decreases but not until an extremely large
sample size is used. The numeric results from the simulation studies with 21 observation
points are shown in Table 2 and confirm the findings described for Figure 2.

The efficiency of the four methods can be evaluated using the Monte Carlo standard errors shown in Table 2.
These standard errors indicate that the
variability of the estimation procedures are quite similar, with the NLS method providing slightly more
precise (lower standard error) estimates.  This is expected, since in this setting, NLS is the maximum likelihood
method for parameter estimation.  In general, the Monte Carlo standard errors for estimates obtained with the PsLS method are slightly larger
than those obtained with BCLS, possibly due to differencing of the data required by methods which rely on differentiation.
The Monte Carlo standard errors from all three methods for the simulations with 201
equally spaced time points between $t=0$ and $t=20$ are show a similar pattern in standard errors from the three methods.
(Results not shown).

\section{Example 2: Nonlinear System of ODEs with two compartments}

The Fitzhugh-Nagumo system of
differential equations \cite{Fit:61,Nag:62} is a simplified version of
the the well known Hodgkin-Hukley model \cite{Hod:52} for the behavior of spike potentials in
the giant axon of squid neurons. The equations
\begin{eqnarray}
 \frac{dV}{dt}= C (V - \frac{V^3}{3} + R), \qquad V(0)=v(0) \label{fhV} \\
 \frac{dR}{dt}= -\frac{1}{C}(V - a + bR),\qquad R(0)=r(0) \label{fhR}
\end{eqnarray}
describe the reciprocal
dependencies of the voltage compartment $V$ across an axon
membrane and a recovery compartment $R$ summarizing outward currents. Solutions to this system
of differential equations quickly converge for a range of starting values to periodic
behavior that alternates between smooth evolution and sharp changes (bifurcations) in direction.
These solutions exhibit features common to elements of biological neural networks
\cite{Wilson:99}.  The system has been used by a number of authors including \cite{RHCC:07,Lia:08}
to demonstrate methods of parameter estimation for data with time varying expectation given by this
system of differential equations.

For this example there are two random variables: the voltage across
an axon membrane, $Y_1({\bf t})$, and a recovery measurement of
outward currents, $Y_2({\bf t})$. We assume that observed data,
$y_1({\bf t})$ and $y_2({\bf t})$ follow independent normal
distributions with
\begin{eqnarray}
 Y_1({\bf t})  \  = \ V({\bf t})\ + \beps_1,\ \beps_1 \   \sim \  i.i.d.\;  N(0,\sigma_1^2) \label{YV} \\
 Y_2({\bf t}) \  = \ R({\bf t})\ + \beps_2,\ \beps_2 \ \sim\  i.i.d.\;  N(0,\sigma_2^2) \label{YR}
\end{eqnarray}
where $V(t)$ and $R(t)$ satisfy the differential equations (\ref{fhV}) and (\ref{fhR}).

The likelihood surface for data with expectation given by a nonlinear system
of differential equations may be extremely complex and can contain multiple local maxima
(equivalently, local minimum solutions to the nonlinear least
squares regression model). For the Fitzhugh-Nagumo equations we
considered two sets of parameters; $C=3$, $a = 0.58$, $b=0.58$ and
$C=3$, $a = 0.34$, $b=0.20$.  The parameters $C=3$, $a = 0.58$,
$b=0.58$ are very near a supercritical Hopf bifurcation value of the system; from a
stable steady state to an oscillating limit cycle.  There is a
dramatic change in the least squares surface as the parameters $a$
and $b$ vary about 0.58 (with $C = 3$ fixed).  Contours of the
least squares surface for a set of zero noise data are shown in
Figure 3 panels A, B,
and C. Panel A shows the contours of the least square surface for a
range of $a$ = 0.4 to 1.4 and $b$ = 0 to 0.2 to 2.2.  The blue box
contains the true solution, $a=0.58$ and $b=0.58$, the red box contains a local minimum.
Panels B and C show close ups of the red and blue regions, respectively.
Panel B suggests that there is local minimum in the least squares fit surface
at approximately $a$ = 1.5, $b$ = 2.0.  Panel C suggests that the least squares
fit surface is badly behaved near the true solution, with a long ridge containing the true
solution, which is near a bifurcation values for the parameters $a$ and $b$.

Figure 3 Panels D,E, and F show similar regions of a least squares surface for zero noise
observations from the Fitzhugh-Nagumo system with parameters $C=3$, $a=0.34$, and
$b=0.2$.  This system has a local minimum shown in the red square and
the true solutions shown in the blue square Panel D.  Close-ups
of these regions are shown in panels E and F respectively.  For this
collection of parameters, the least squares surface near the true solution
is well behaved (Panel F), however, a local minimum in the surface
exists near $a = 1.7$, $b$ =2.8 (Panel E).

\subsection{Simulations}

To compare the BCLS method to NLS and PsLS for accuracy and efficiency in estimating
parameters in the Fitzhugh-Nagumo model, data was simulated from the
statistical model (\ref{YV}) and (\ref{YR}), for two sets of parameter values; $C$ =3,
$a$ = 0.34, $b$ = 0.2 and $C$ =3, $a$ = 0.58, $b$ = 0.58.  For each of the
two sets of parameter values, a range of residual standard errors $\sigma_1$ and
$\sigma_2$ were evaluted ; $\sigma_1$ = 0.05, 0.1, and 0.15, and
$\sigma_2$ = 0.05, 0.1, and 0.15.
Time points evenly spaced time points at intervals of 0.1 between $0$ and $20$ for a total of 201 time points were considered.
Initial conditions for the system,
$v_0$ = -1 and $r_0$ =1 were used for all simulations. The data were
simulated 1000 times for each set of parameter values and
values of $\sigma_1$ and $\sigma_2$. All three parameters
$C$, $a$, and $b$ as well as the initial conditions, $v_0$ and $r_0$ were estimated
using nonlinear least squares, bias-corrected least squares and least square without bias correction.
Since PsLS relies on derivatives rather than integrals, estimation of the intial conditions $v_0$ and $r_0$ is not possible,
so that only estimates of the three model parameters, $a$, $b$ and $C$ were obtained using the PsLS method.

For each of the 1000 replicates, nonlinear least squares regression
of all five values is conducted simultaneously
using simulated data $y_1({\bf t})$ and $y_2({\bf t})$ and solutions to
(\ref{fhV}) and (\ref{fhR}) to form the weighted (when $\sigma_1^2 \neq \sigma_2^2)$ least squares expression
\[ \sum_{i=1}^n \frac{(y_1(t_i) - V(t_i))^2}{\sigma_1^2} + \frac{(y_2(t_i) - R(t_i))^2}{\sigma_2^2}. \]
The BCLS estimation is conducted in two steps. First,the parameter $C$  and initial condition $v_0$ are estimated using
data $y_1({\bf t})$ a regression model based on (\ref{fhV}).  Next the parameters $a$ and $b$ and initial condition $r_0$ are estimated using data $y_2({\bf t})$, a regression model based on (\ref{fhR}), and the estimate of $C$ obtained in the first step.

To estimate $C$ and $v_0$, equation (\ref{fhV}) is re-written as the integral equation
\begin{eqnarray}
V(t)  & = &  C \int_0^{t} h_{1,1} (V,R) d\tau  + v_0 \label {fhintV}
\end{eqnarray}
where $h_{1,1}(V,R) = V - \frac{V^3}{3} + R$. Since $E[Y_1^3] \neq E[Y_1]^3$ a bias correction function
must be identified.  Calculating $E[Y_1-\frac{Y_1^3}{3}+Y_2] = V + (\frac{V^3}{3} + 3 V \sigma_1^2) + R$,
it is clear that
\[h_{1,1}^{\star}(V,R) = V - \frac{V^3}{3} + R -  3 V \sigma_1^2\]
satisfies condition \ref{ehstar}.
Therefore, to approximate the
integral in (\ref{fhintV}) we
define the covariates $z({\bf t})=\{z(t_0),z(t_1),\cdots,z(t_n)\}$ using the trapezoid rule for integral approximation:
$z(t_0)=0$ and for $i=1,\cdots,n$
\begin{eqnarray*}
 z(t_i) & = &  \sum_{j=1}^{i} \frac{ \left(h_{1,1}^{\star}\left[y_1(t_j),y_2(t_j)\right] +  h_{1,1}^{\star}\left[y_1(t_{j-1}),y_2(t_{j-1})\right] \right) *(t_j-t_{j-1})}{2}
\end{eqnarray*}
Finally, since  the initial value, $v_0$, will be estimated, the response
for the regression model for the BCLS method estimates of $v_0$ and $C$ is given by $y'({\bf t})=y_1({\bf t})$.
To obtain the integrated-data estimates of  $v_0$ and $C$ we
fit a linear regression model with intercept based on
\begin{eqnarray}
y'({\bf t}) \sim v_0\ {\bf 1}+ C\ z({\bf t}) + \beps_1,
\quad  \beps_1 \sim i.i.d. \;  N(0,\sigma_1^2 ).
\label{FN1regmod}
\end{eqnarray}
The estimated coefficient of $z({\bf t})$ will provide an unbiased estimate of $C$, denoted by ${\hat C}$ and the e
stimated intercept will provide an unbiased estimated of $v_0$. When $h_{1,1}$ is used to create the
covariates, rather than $h^{\star}_{1,1}$, biased estimates of $C$ and $v_0$ are expected.

Using the estimated value ${\hat C}$ the BCLS method estimates of $a$, $b$ and $r_0$  are obtained
by first transforming (\ref{fhR}) to the integral equation
\begin{eqnarray}
{\hat C}\ R(t) + \int_0^{t} V(\tau) d\tau &=& r_0 + a {\bf t} - b \int_0^{t} R(\tau) d\tau \label{fhintR}.
\end{eqnarray}
In this case integral approximations using the observed data are required to define both the response and a covariate
for the linear regression model that will be used to estimate $a$, $b$ and $r_0$.
The response for the linear regression model approximates the left-hand side of (\ref{fhintR})
and is denoted by $y'({\bf t}) =  \{ y'(t_0),y'(t_1),\cdots,y'(t_n)\} $ where $y'(t_i)$ is defined as
$y'(t_0) = {\hat C} y_2(t_0)$ and for $i=1,\cdots,n$:
$$
y'(t_i) =  {\hat C}\  y_2(t_i)) +  \sum_{j=1}^{i} \frac{ \left[ y_1(t_j) +  y_1(t_{j-1}) \right] *(t_j-t_{j-1})}{2}.
$$
The covariates for
the linear regression model are ${\bf t}$ and $z({\bf t})$ where $z({\bf t})$ approximates $\int_0^{t} R(\tau) d\tau$ and
is calculated using the observed data, $y_2({\bf t})$, and the trapezoid rule.  There is no need for bias
correction since $E[Y_2] = R$.  The covariates are defined as $z(t_0)=0$ and for $i=1,\cdots,n$
\begin{eqnarray*}
 z(t_i) & = &  \sum_{j=1}^{i} \frac{ \left[ y_2(t_j) +  y_2(t_{j-1}) \right] *(t_j-t_{j-1})}{2}
\end{eqnarray*}

To obtain estimates of $a$, $b$ and $r_0$ the following linear regression model with intercept is used.
\begin{eqnarray*}
y'({\bf t}) \sim r_0\ {\bf 1} + a\ {\bf t}- b\ z({\bf t}) + \beps,
\qquad  \beps \sim i.i.d. \;  N(0,\sigma^2 ).
\end{eqnarray*}
The estimated intercept provides an estimate of $r_0$, the estimated coefficient of ${\bf t}$ provides an estimate of $a$
and the estimated coefficient of $z({\bf t})$ provides an estimate of $b$.

A similar two-stage approach was used to obtain estimates of $a$, $b$, and $C$ using the PsLS method. Since this method
relies on derivatives rather than integrals, estimation of the intial conditions $v0$ and $r_0$ is not possible.
The PsLS method estimates using local polynomial smoothing with bandwidth of order $\log(nt)^{-1/2}*nt^{-2/7}$ (0.20).
A variety of other bandwidths were considered before choosing the one which resulted in the smallest overall bias in parameter estimates.
Simulations were also conducted using PsLS with a smoothing bandwidth of order $\log(nt)^{-1/10}*nt^{-2/7}$ (0.37)
to evaluate sensitivity of this method to choice of bandwidth.

The bias from each method as a function of measurement error is shown in Figure 4.
Table 3 provides the Monte Carlo means and standard errors for LS, BCLS, and PsLS estimated parameters based on simulations
equal values for $\sigma_1 = \sigma_2 = 0.05$. The simulation results
with all combinations of $\sigma_1$ and $\sigma_2$ provided similar results and are not shown. Nonlinear least squares estimates
were unbiased with the lowest Monte Carlo standard errors (results not shown). However, NLS failed to converge in 7\% of the simulations
with $a=0.34$ and $b=0.2$ and in 9\% of the simulation with  $a=0.58$ and $b=0.58$ when the true values were used as initial conditions
for the NLS algorithm.
As expected, bias was detected in the LS estimated parameters when bias correction was not used.
Most affected were the estimates of $C$ , $v_0$, and $r_0$.  Using the
BCLS method resulted in elimination of most of the bias observed in the unadjusted LS estimated parameters.  The PsLS method
with a bandwidth of 0.2 produced less biased and more accurate estimates of the parameter $C$, but slightly more biased and less accurate
estimates of the parameters $a$ and $b$.  When bandwidth of 0.37 was used, the PsLS method produced estimates with more bias than the
BCLS method.

Since choice of starting values can influence NLS parameter estimation, additional analysis of simulated data
were performed with the incorrect value of the parameters used for starting estimates in the NLS algorithm.
For these analysis, data
from the single compartment for recovery of outward current, $Y_2({\bf t})$, were used as outcomes in the nonlinear least square regression analysis
and only the parameters $a$ and $b$ estimated.
The results of these analysis for simulated data with 201 equally spaced time points and residual standard error, $\sigma_2= 0.05$
are shown in Table 4 for the model with $a=0.34$ and $b=0.2$ (first 3 columns) and for the model with $a=0.58$ and $b=0.58$ (last 3 columns).
The values $a=0.4,0.8$, and $1.2$ and $b=0.4$ and $0.8$
were used as starting values for the NLS algorithm. The BCLS estimates of $a$ and $b$ were also used as starting values
for the NLS algorithm.

In most cases, when the incorrect values of the parameters were used as starting values for NLS regression, the algorithm
failed to converge, and when convergence occurred, the estimated values of the parameters were incorrect.  For example,
with the true value of $a = 0.34$ and $b=0.2$ if starting values of $a=0.8$ or $a=1.2$ is used (with either $b=0.4$ or $b=0.8$)
the algorithm converged for approximately 35\% of the simulation replicates.  When the algorithm did converge, the Monte Carlo
average of the estimate of $a$ is approximately $1.72$ and the average estimate of $b$ is approximately $2.77$.  These values
correspond the local minimum of the least squares surface shown in Figure 3 Panel E. Similarly, when the true value of $a$ and $b$
are both $0.58$, for most of the incorrect starting estimates, the NLS algorithm converged less for less than 35\% of the simulations
The Monte Carlo average of the estimate of $a$ when the NLS algorithm did converge to the wrong value is approximately $1.15$ and the average estimate of
$b$ is approximately $2.02$ which corresponds to the local minimum in the least squares surface for those parameters
(Figure 3 Panel B). When BCLS method was used to obtain starting values for the NLS regression algorithm, unbiased estimates of both $a$ and $b$
are obtained, and the NLS algorithm converged for more than 90\% of the simulation replicates.

\section{Discussion}

The bias-corrected least squares method is a computationally simple, non-iterative, and easily implemented
method for estimation of parameters in ODE models.  It's development and the associated proof of
consistency of the resulting parameter estimates illustrates that direct methods (non-iterative) can be modified to
so that the resulting estimator has desirable statistical properties.  The BCLS method retains the simplicity of early direct
methods and has most of the desirable statistical properties of more recently
developed direct methods. A key advantage of the BCLS method (and all direct methods) is that it
does not require starting values for the estimation algorithm; in fact, it can provide starting values for
nonlinear least squares regression if desired. Furthermore, it does not require choice of smoothing bandwidth of functional data analysis
methods.  Since the BCLS method is based on
integration rather than differentiation it can be used to estimate initial states of the ODE model
states in addition to ODE model parameters. This is not true of direct methods which rely on differentiation such as PsLS.

The analysis and simulations described in Section 3 based on the population growth of paramecium aurelium
show that the BCLS method performs comparably with the NLS method.  In the simulation
studies conducted using the logistic growth equation, the BCLS method produces estimates that are often less biased
than the recently developed PsLS method, a direct method which uses differentiation and smoothing instead of integration.
The bias observed in the PsLS method is likely a result of the choice of bandwidth in the smoothing step;
the use of integration in the BCLS method alleviates the need for any smoothing of the data since integration
itself is a smoothing procedure. However, the bias due to integral approximations required by the BCLS can be expected when the
data are not densely sampled. The simulation studies conducted with the Fitzhugh Nagamu equations in Section 4
provide further evidence that the BCLS method
is a viable alternative to NLS or PsLS regression.  Those simulations demonstrate that
the use of the BCLS method to obtain starting values for NLS regression can reduce problems
with convergence and accuracy associated with the choice of starting values in the NLS method,
especially in settings with an exceptionally complex likelihood surface or cost function with
multiple minima.  These simulations also show that while the PsLS method has the capacity to produce less biased
and more accurate estimates for some parameters, these estimates rely heavily on the choice of bandwidth.
Furthermore, since BCLS is based on integration and PsLS is based on differentiation, the BCLS method is
able to provide estimates of the initial states of a system of ODE's. This not possible with the PsLS method.
While both the BCLS method and the PsLS method are asymptotically unbiased, the simulations in this paper suggest that
the BCLS method has better small sample properties if an optimal bandwidth is not used with PsLS. Furthermore, simulations
suggest that a bandwidth which produces the least bias in one estimated parameter may not be the optimal choice for all
parameters in the system.

There are a number of drawbacks to the BCLS method.  In order to obtain unbiased
parameter estimates a separate analysis may be needed to obtain estimates of residual
errors to be used in the bias correction functions.  Although the linear regression
algorithm is used to obtain point estimates for the parameters, this approach is not
likelihood based so regression techniques for conducting inference or obtaining
confidence intervals cannot be used directly. However, bootstrapping techniques are
available to obtain confidence intervals for the estimated parameters. Another drawback
to the BCLS method is the assumption that parameters must appear linearly in the ODE
system (A.1).  This excludes estimation of certain parameters in models such as the
Michaelis-Menten model for enzyme kinetics or saturating growth models for population
dynamics.  Nonlinear estimation techniques could be used when parameters appear
nonlinearly in the equations motivated by Eq (\ref{ey}), and may offer improvements over
directly solving the ODE and applying nonlinear estimation techniques to the resulting
solution.  This is an approach that was not considered in this work.  However, when known
parameter appears nonlinearly in the system of ODEs it can be used to create the
covariates (or response) for the bias-corrected least squares method as demonstrated in
the estimation of $a$ and $b$ in the Fitzhugh-Nagumo model in Section 4.

In conclusion, the bias-corrected least squares method offers a relatively straightforward
and easily implemented consistent estimation technique for parameters and initial conditions in
models with means defined by ordinary differential equations. A variety of extensions of the methods
are being evaluated including relaxing the assumptions that all states be observed at the same time points
as well as evaluating the statistical properties of other estimation procedures which improve on the original direct methods
proposed by Himmelbau, Jones and Bischoff (HJB). The identification of the bias corrections function in this work
suggests that there may be relatively straightforward modifications that can be made to a variety of direct methods so that
the resulting estimates have desirable statistical properties. Simple modifications and the study of associated statistical
properties of direct methods similar to the HJB methods is a promising area that is likely to provide new tools to investigators
working with data generated by mechanisms that can be described with ordinary differential equations.


\section{Appendix}

\subsection*{Proof of Theorem 1}

Recall that upper case notation, $Y$ and $Z$
indicates that we are working with random variables rather than observed data,
which we represent with lower case variables.
The bias-corrected least squares method (BCLS) is appropriate for random variables
${\bf Y}({\bf t})=(Y_1({\bf t}),\cdots,Y_m({\bf t}))$ which are
observed at times ${\bf t}=(t_0,\cdots,t_n)$ and have time varying expected
values given by the compartments (states) of a system of differential equations:
  \begin{eqnarray}
  \frac{d\mu_q}{dt}=\sum_{k=1}^{m_q} \beta_{q,k} h_{q,k}(\bmu), \ \
  \mu_q(t_0)=\mu_{q,0}, \ \ q=1,\cdots,s \ \
  \end{eqnarray}
Specifically, the expectation, $E\{{\bf Y}({\bf t})\}$ is equal to $ \bmu({\bf t})$
and the conditions (A.1)-(A.3) listed in Section 2 are satisfied. To simplify notation
notation, we assume that the interval lengths defined by the observation
times, $(t_{i+1}-t_i)$ are equal to $\frac{1}{n}$. The
results hold for any choice of sampling times where the maximum
interval length defined by the sampling times is $O(n^{-1})$.

As described in Section 2, we define covariates  $Z_{q,k}({\bf t})$, $q=1,\cdots s$, $k=1\cdots m_q$,
using the left endpoint integral approximation as:
\begin{eqnarray*}
Z_{q,k}(t_0) & = & 0 \\
Z_{q,k}(t_i) & = & \sum_{j < i} h_{q,k}^{\star}\{ {\bf Y}(t_j) \} \frac{1}{n}, \;\; i=1,\cdots, n
\label{covdefn}
\end{eqnarray*}
where the functions $ h_{q,k}^{\star} $ are described in (\ref{ehstar}). Note that as functions of the
random variables ${\bf Y}(t_j)$ these covariates are also random variables.

To further ease of notation, we now drop the subscript $q$ and describe results for a single regression analysis
using data, $\{y_1,\cdots,y_n\}$, with expectation from an ODE with a single compartment.
The results are easily extended
for data with expectations from a system of ODE's with more than one compartment.
If ${\bf z }({\bf t})$ is the covariate matrix with columns $z_{k}({\bf t}), \ k=1,\cdots,m$,
then the parameters $\Vb = \{\beta_1,\cdots,\beta_m\}$ are estimated using linear regression so that
\[
\wVb =  ({\bf z }({\bf t})'{\bf z }({\bf t}))^{-1}{\bf z}({\bf t})'{\bf z}({\bf t})
\]
Using results from \cite{Cro:86} consistency of $\wVb$ is achieved by showing
that if  $\wVb_n$ is the unique root of the estimating function  $\displaystyle \U_{n}(\Vb)=\frac{1}{n}
\sum_{i=1}^n \bz_i^T  \left\{ \by_i - \bz_i  \Vb \right\}=0$ and Conditions (A1) - (A3) are satisfied then:
\begin{itemize}
\item[I]  $\U_{n}(\wVb_n)$ is a consistent estimator of zero.
\item[II] $E\{\frac{1}{n} \sum_{i=1}^n \bZ_i^T \bZ_i \}$ is bounded and
non-zero.
\item[III]  $ \mbox{var}\{\frac{1}{n} \sum_{i=1}^n \bZ_i^T \bZ_i\} \rightarrow 0$  as $n \rightarrow \infty$.
\end{itemize}

\subsection*{I\ \  $\U_{n}(\wVb_n)$ is a consistent estimator of zero}

To establish that  $\U_n(\Vb)$ is a consistent estimator of zero  we need to show that,
\begin{eqnarray*}
\lim_{n\rightarrow\infty} E [ \U_n(\Vb) ] & = & 0 \\
\lim_{n\rightarrow\infty} \mbox{var} [ \U_n(\Vb) ] & = &  0
\end{eqnarray*}
We consider the situation where $\Z({\bf t})=Z({\bf t)}$ is a column vector, i.e., there
is a single covariate for the regression problem so that $\Vb$ is a scalar, $\beta$ Since
$Y(t_1),\cdots,Y(t_{i-1}),Y(t_i)$ are independent and the covariates $Z(t_i)$ are based
only first $i-1$ of these random variables (using the left endpoint approximation for
integrals), it follows that $Z(t_i)$ and $Y(t_i)$ are independent so that $E\{ Z(t_i)
Y(t_i) \} = E\{Z(t_i)\} ; E\{Y(t_i)\}$

To further simplify the notation define:
$Z(t_i)  =  Z_i ,\
Y(t_i)  =  Y_i ,\
E\{ Z(t_i) \}  =  M_i ,\
E\{ Y(t_i) \}  =  \mu_i ,\
U_i  =  Z_i ( Y_i - Z_i \beta ) ,\
Z_{i \backslash j}  =  Z_i-Z_j = \frac{1}{n} \displaystyle{\sum_{j\leq l < i}} h^{\star}(Y_l).
$
To study the limit properties of $\U_n(\beta)$ the following results will be useful:
\begin{eqnarray}
E(Z_i)  &  = & O(n^{-1}) \; O(i)  \label{r0} \\
\mbox{var}(Z_i) & =  & O(n^{-2}) \; O(i)  \label{r1} \\
\mbox{cov}(Z_i,Z_i^2) & = & O(n^{-3}) \; O(i^2) \label{r2}  \\
\mbox{var}(Z_i^2) & = & O(n^{-4}) \; O(i^3)  \label{r3} \\
\mbox{for } j<i,\ \mbox{cov}(Z_j^2,Z_i^2) & = & O(n^{-3})\ O(j^2) + O(n^{-4})\ O(j^3) \label{r4}
\end{eqnarray}

It is straightforward to see that $E(Z_i)$ and $\mbox{var}( Z_i )$ are
bounded. This follows from (A.3) and
\[ E(Z_i)\ =\   \frac{1}{n} \displaystyle{\sum_{j<i}} h^{\star}(Y_j)\  =
\  \frac{1}{n} \displaystyle{\sum_{j<i}} h\{\mu(t_j)\}\  \leq \ \frac{i}{n} B \  \leq\   B \]
and
\[ \mbox{var}( Z_i )\  =\  \mbox{var} \left\{
   \frac{1}{n} \sum_{j < i} h^{\star}( Y_j ) \right\} \
   = \  \frac{1}{n^2} \sum_{j < i} \mbox{var}\{ h^{\star}( Y_j ) \} \
   \leq \  \frac{i}{n^2} B \  \leq \   B \]

To show that $E\{ \U_n(\beta) \} \rightarrow 0$ as $n \rightarrow \infty$ we calculate
\begin{eqnarray*}
E\{ \U_n(\beta) \} & = & \frac{1}{n} \sum_{i=1}^n E(Z_i \; Y_i) - E(Z_i^2)\; \beta \\
  & = & \frac{1}{n} \sum_{i=1}^n E(Z_i) \; E(Y_i) \; - \; E( Z_i)^2 \; \beta - \;  \mbox{var}( Z_i ) \; \beta\\
  & = & \frac{1}{n} \sum_{i=1}^n E(Z_i) \left[ \beta \int_{0}^{t_i}
  h\{\mu(s)\} ds -\beta \sum_{j<i} h\{\mu(t_j)\} \frac{1}{n} \right] - \;  \mbox{var}( Z_i ) \; \beta \\
  & = & \frac{1}{n} \sum_{i=1}^n E(Z_i)\; \sum_{j<i} \beta \left[  \int_{t_{j-1}}^{t_{j}}
  h\{\mu(s)\} ds -  h\{\mu(t_j)\} \frac{1}{n} \right] - \; \mbox{var}( Z_i ) \;  \beta \\
  & = & E_n^{(1)}(\beta) \;\; + \;\; E_n^{(2)}(\beta ) \\
\mbox{where} & & \\
 E_n^{(1)}(\beta) & = & \frac{1}{n} \sum_{i=1}^n E(Z_i)\; \beta\;
    \sum_{j<i} \left[ \int_{t_j}^{t_{j+1}} h\{ \mu(s)\}- h\{\mu(t_j)\} ds \right] \\
 E_n^{(2)}(\beta) & = & \frac{1}{n} \sum_{i=1}^n \mbox{var}( Z_i )\; \beta
\end{eqnarray*}

Let $\epsilon_n =
 \mbox{max }_{i=1,\cdots, n} \left[\  |\ h\{ \mu(s)\}-
 h\{\mu(t_i)\}\ |\  :\  s\in [t_{i-1},i_j]\  \right\}$.
Since $h$ and $\mu$ are continuous on $[0,t_n]$ it follows
that $\epsilon_n \rightarrow 0$ as $n\rightarrow \infty$ and
$$\int_{t_j}^{t_{j+1}} |\ h\{ \mu(s)\}- h\{\mu(t_j)\}|\  ds \leq
\epsilon_n \frac{1}{n}.$$ Since $E(Z_i)$ is bounded for all $i$,
we have,
\[
E_n^{(1)}(\beta) \   \leq  \   \frac{B\beta}{n} \sum_{i=1}^n \sum_{j<i}\frac{\epsilon_n}{n}
 \  = \   O(n^{-2})\  \epsilon_n\  \sum_{i=1}^n i
 \  = \   O(n^{-2})\  O(n^2)\  \epsilon_n
 \  \rightarrow \  0  \mbox{ as }  n\rightarrow \infty
\]
We use variance bounds to show that $E_n^{(2)}(\beta)\rightarrow 0$ as $n\rightarrow \infty$.
\[
E_n^{(2)}(\beta) \  = \  \frac{1}{n} \sum_{i=1}^n \mbox{var}( Z_i ) \beta
  \  = \  O(n^{-1})\  \sum_{i=1}^n O(n^{-2})\ O(i)
  \  = \   O(n^{-3}) \ O(n^2)  \  \rightarrow \  0 \mbox{ as }  n\rightarrow \infty
  \]
This completes the proof that $\lim_{n\rightarrow\infty} E [ \U_n(\Vb) ]  =  0$.
\vskip0.1in

Next, we consider $\mbox{var}\{\U_n(\beta)\}\  =\  \frac{1}{n^2} \left\{ \sum_{i=1}^n\mbox{var}(U_i)
 + 2 \sum_{j<i} \mbox{cov}(U_j,U_i) \right\}$.
First we evaluate $ \mbox{var}(U_i) =  \mbox{var}(  Z_i Y_i) + \mbox{var}(  Z_i^2 \beta)
- 2 \mbox{cov}(Z_i Y_i,Z_i^2 \beta) $. Using equations (\ref{r0})-(\ref{r4})
and independence
of $Z_i$ and $Y_i$ we obtain
\begin{eqnarray*}
 \mbox{var}(  Z_i\ Y_i) & = & E[Z_i^2\ Y_i^2] - E[Z_i\ Y_i]^2 \\
   & = &  E[Z_i^2]\ E[Y_i^2] - E[Z_i]^2\ E[Y_i]^2 \\
   & = &  (\mbox{var}(Z_i)-M_i^2)\ (\mbox{var}(Y_i) - \mu_i^2) - M_i^2 \mu_i^2 \\
   & = &  \{\mbox{var}(Z_i) \; \mbox{var}(Y_i) \} - \mu_i^2 \mbox{ var}
   ( Z_i) - M_i^2 \mbox{ var} (Y_i) \\
   & = &  O(n^{-2})\  O(i) - O(n^{-2})\ O(i^2) \\
\mbox{var}( Z_i^2 \beta) & = & O(n^{-4})\  O(i^3) \\
\mbox{cov}( Z_i\ Y_i \; Z_i^2 \beta) & = & \beta\ \mu_i
    \mbox{ cov}(Z_i,Z_i^2) \\
    & = & O(n^{-3}) \; O(i^2)
\end{eqnarray*}

Thus, we have a contribution to $\mbox{var}[\U_n(\beta)]$ from the
diagonal (variance) terms of:
\begin{eqnarray*}
\frac{1}{n^2} \sum_{i=1}^n \mbox{var}(U_i) & = & O(n^{-2})\
 \sum_{i=1}^n\ \left\{  O(n^{-2})\ O(i) + O(n^{-2})\ O(i^2) +  O(n^{-4})\ O(i^3) + O(n^{-3}) \;
 O(i^2) \right\} \\
 & = &  O(n^{-2}) \left\{  O(n^{-2})\  O(n^2) + O(n^{-2})\ O(n^3)\ +\ O(n^{-4})\  O(n^4) + O(n^{-3}) \ O(n^3) \right\}\\
 & = &  O(n^{-2}) + O(n^{-1}) \ \rightarrow \ 0 \mbox{ as }
 n\rightarrow \infty\end{eqnarray*}
To evaluate $ \mbox{cov}(U_j,U_i)$ assuming $j<i$ consider:
\begin{eqnarray*}
 \mbox{cov}(U_j,U_i) & = &  \mbox{cov}( Z_j Y_j-Z_j^2\beta,Z_iY_i - Z_i^2\beta ) \\
  & = &  \mbox{cov}( Z_jY_j,  Z_iY_i  )   -
   \mbox{cov}( Z_jY_j,  Z_i^2\beta  )  -
   \mbox{cov}(  Z_j^2\beta,   Z_iY_i  )   +
   \mbox{cov}( Z_j^2\beta,   Z_i^2\beta  )
\end{eqnarray*}

Using equations (\ref{r0})-(\ref{r4}) it follows that:
\begin{eqnarray*}
  \mbox{cov}(Z_j Y_j, \; Z_iY_i ) & = &  \mbox{cov}(Z_j Y_j,(Z_{i\backslash (j+1)} + Z_{j+1}) Y_i) \\
                                  & = &  \mbox{cov}(Z_jY_j,Z_{i\backslash (j+1)} Y_i) + \mbox{cov} (Z_j Y_j,Z_{j+1} Y_i) \\
                                  & = &  0\; + \mbox{cov} (Z_j Y_j, [\frac{1}{n}h^{\star}(Y_j) + Z_j ] Y_i)  \\
                                  & = &  \frac{1}{n} \{\mbox{cov} (Z_j Y_j, h^{\star}(Y_j) Y_i)\} +\mbox{cov}(Z_j Y_j,Z_j Y_i)\\
                                  & = &  \frac{1}{n} \{ E[Z_j Y_j h^{\star}(Y_j) Y_i] - M_j \mu_j h^{\star}(\mu_j) \mu_i  \}
                                                + \mu_j \mu_i \mbox{var}(Z_j) \\
                                  & = &  \frac{1}{n} \{ M_j \mu_j E[Y_j h^{\star}(Y_j)] - M_j \mu_j h^{\star}(\mu_j) \mu_i  \}
                                                + \mu_j \mu_i \mbox{var}(Z_j) \\
                                  & = &  \frac{1}{n} \{ M_j \mu_j \left(E[Y_j h^{\star}(Y_j)] - h^{\star}(\mu_j) \mu_i \right) \}
                                                + \mu_j \mu_i \mbox{var}(Z_j) \\
                                  & = &   O(n^{-2}) \;\; O(j) \\
\end{eqnarray*}
Similar calculations show that
\begin{eqnarray*}
\mbox{cov}( \; Z_j Y_j, \; Z_i^2 \beta \; ) & = &  O(n^{-2}) \;\; O(j) + O(n^{-3}) \;\; O(j^2) \\
\mbox{cov}( \; Z_j^2\beta, \; Z_iY_i \; ) & = & O(n^{-3}) \;\; O(j^2) \\
\mbox{cov}( \; Z_j^2 \beta, \; Z_i^2 \beta \; ) & = &  O(n^{-3}) \;\; O(j^2) + O(n^{-4}) \;\; O(j^3)
\end{eqnarray*}

It follows that the contribution to $\mbox{var}\{\U_n(\beta)\}$ from the
off-diagonal terms is:
\begin{eqnarray*}
2 \frac{1}{n^2} \sum_{i=1}^n \sum_{j<i} \mbox{cov}(U_i,U_j)  & = &
    O(n^{-2}) \sum_{i=1}^n \sum_{j<i} \; O(n^{-2}) \  O(j)  \\
  &  &  \; - \  O(n^{-2})  \sum_{i=1}^n \sum_{j<i}  O(n^{-2}) \  O(j) + O(n^{-3}) \  O(j^2) \\
  &  & \; - \   O(n^{-2})  \sum_{i=1}^n \sum_{j<i}  O(n^{-3}) \  O(j^2) \\
  &  &  \; + \  O(n^{-2})  \sum_{i=1}^n \sum_{j<i}  O(n^{-3}) \  O(j^2) + O(n^{-4}) \  O(j^3) \\
  & = & O(n^{-2}) \; O(n^{-2}) \  O(n^3) \\
  &  &  - O(n^{-2}) \ \left\{ O(n^{-2}) \  O(n^3) + O(n^{-3}) \
    O(n^4) \right\} \\
  &  &  - \  O(n^{-2}) \ \left\{ O(n^{-3}) \  O(n^4) \right\} \\
  &  &   + \  O(n^{-2}) \ \left\{ O(n^{-3}) \  O(n^4) + O(n^{-4})
    \  O(n^5) \right\} \\
  & = & O(n^{-1}) \rightarrow 0 \mbox{ as } n\rightarrow \infty
\end{eqnarray*}

This completes the proof that $\lim_{n\rightarrow\infty} \mbox{var} [ \U_n(\Vb) ]  =   0 $.
Combined with the previous proof that $\lim_{n\rightarrow\infty} E [ \U_n(\Vb) ]  =  0$ we
have established that $\U_n(\Vb)$ is a consistent estimator of $0$.

\subsection*{II\ \  $E\{\frac{1}{n} \sum_{i=1}^n \bZ_i^T \bZ_i \}$ is bounded and non-zero}

To see that   $ E \left(\frac{1}{n} \sum_{i=1}^n Z_i^T Z_i \right)$ is bounded and non-zero
we calculate:
\begin{eqnarray*}
 E \left( \frac{1}{n} \sum_{i=1}^n Z_i^T Z_i \right)  & = & \frac{1}{n} \sum_{i=1}^n \{ \mbox{var}(Z_i) + E(Z_i)^2\}  \\
   & = & O(n^{-1}) \left\{ \sum_{i=1}^n O(n^{-2}) \; O(i) + M_i^2 \right\} \\
   & \leq & O(n^{-1}) \; \{O(n^{-2}) O(n^2) + O(n) \} \\
   & = & O(1)
\end {eqnarray*}
so that  $ E\left\{\frac{1}{n} \sum_{i=1}^n Z_i^2\right\}$ is bounded.  As long as
$E(Z_i^2) > 0$ for at least one $i$ we have
$$ E\left(\frac{1}{n} \sum_{i=1}^n Z_i^T Z_i\right) > 0 $$
which holds as long as the original system in not at equilibrium throughout the entire
observation time, which is assumption (A.3).

\subsection*{III\ \  $ \mbox{var}\{\frac{1}{n} \sum_{i=1}^n \bZ_i^T \bZ_i\} \rightarrow 0$  as $n \rightarrow \infty$}

Finally we show that
$\mbox{var}\left(\frac{1}{n} \sum_{i=1}^n Z_i^2\right)
\rightarrow 0 \mbox{ as } n\rightarrow \infty$
by calculating
\begin{eqnarray*}
\mbox{var}\left(\frac{1}{n} \sum_{i=1}^n Z_i^2\right) & = &
    O(n^{-2}) \sum_{i=1}^n \left\{\mbox{var}(Z_i^2) + 2\sum_{j<i}
    \mbox{cov}(Z_j^2,Z_i^2) \right\} \\
 & = & O(n^{-2})\sum_{i=1}^n\left\{ O(n^{-2})\ O(i) + \sum_{j<i}
    O(n^{-2})\ O(j) + O(n^{-4})\ O(j^2) \right\} \\
 & = & O(n^{-2})\sum_{i=1}^n \left\{ O(n^{-2})\ O(i) + O(n^{-2})\
    O(i^2) + O(n^{-4})\ O(i^3) \right\} \\
 & = & O(n^{-2}) \left\{ O(n^{-2})\ O(n^2) + O(n^{-2})\
    O(n^3) + O(n^{-4})\ O(n^4) \right\} \\
 & = & O(n^{-1}) \rightarrow 0 \mbox{ as } n\rightarrow \infty
\end{eqnarray*}

\pagebreak

\begin{figure} \label{fig1}

\caption{BCLS and NLS estimates and associated trajectories based on
data from four growth colonies of paramecium aurelium }
\centerline{
  \epsfig{file=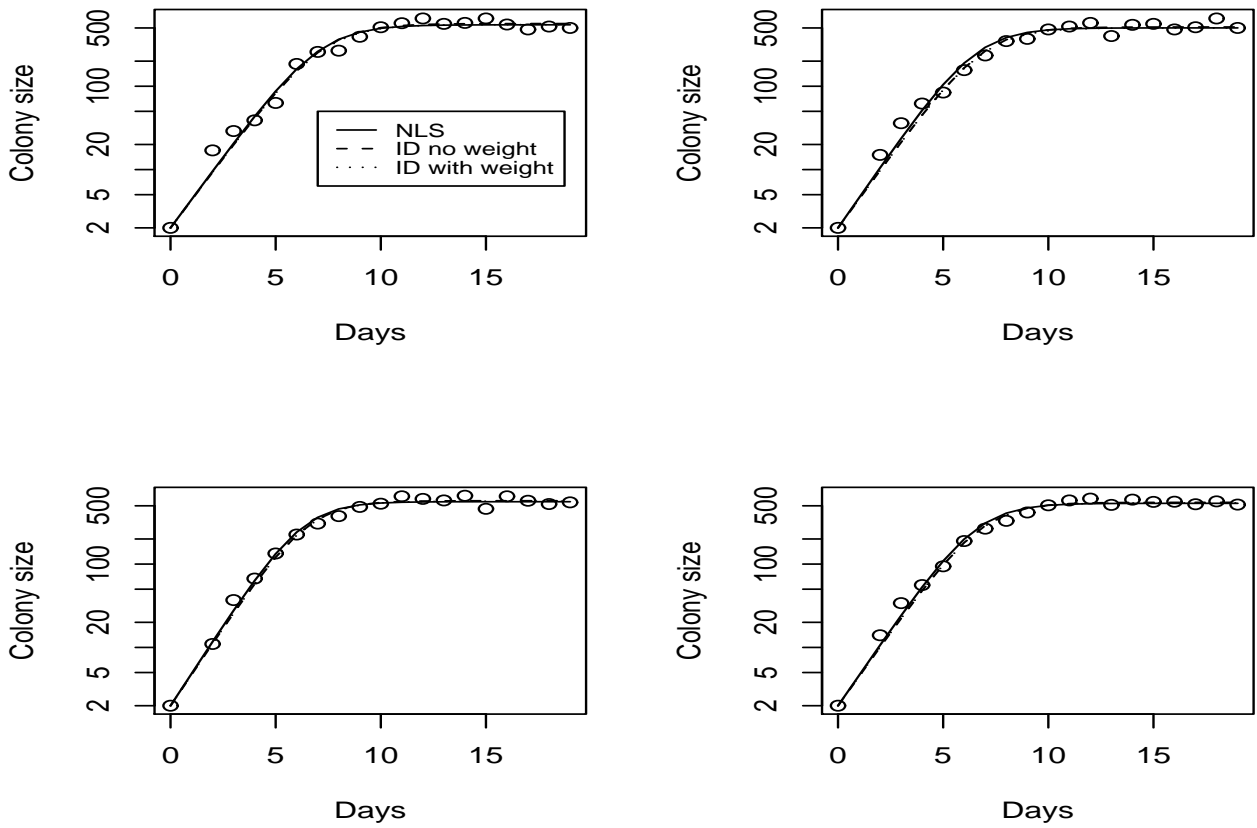,height=5in,width=7in}
}
\end{figure}

\begin{figure} \label{fig2}

\caption{LS, BCLS, NLS and PsLS estimates and associated bias in estimated
parameters based on 1000 simulations with levels of
measurement error varying from 0.2 to 0.8.
The upper panels, A. and B., represent simulations
when 21 times points, evenly spaced between 0 and 20, are
used in the simulations and the lower panels, C. and D., when 201
time points, evenly spaced between 0 and 20, are used in the
simulation.}
\centerline{
  \epsfig{file=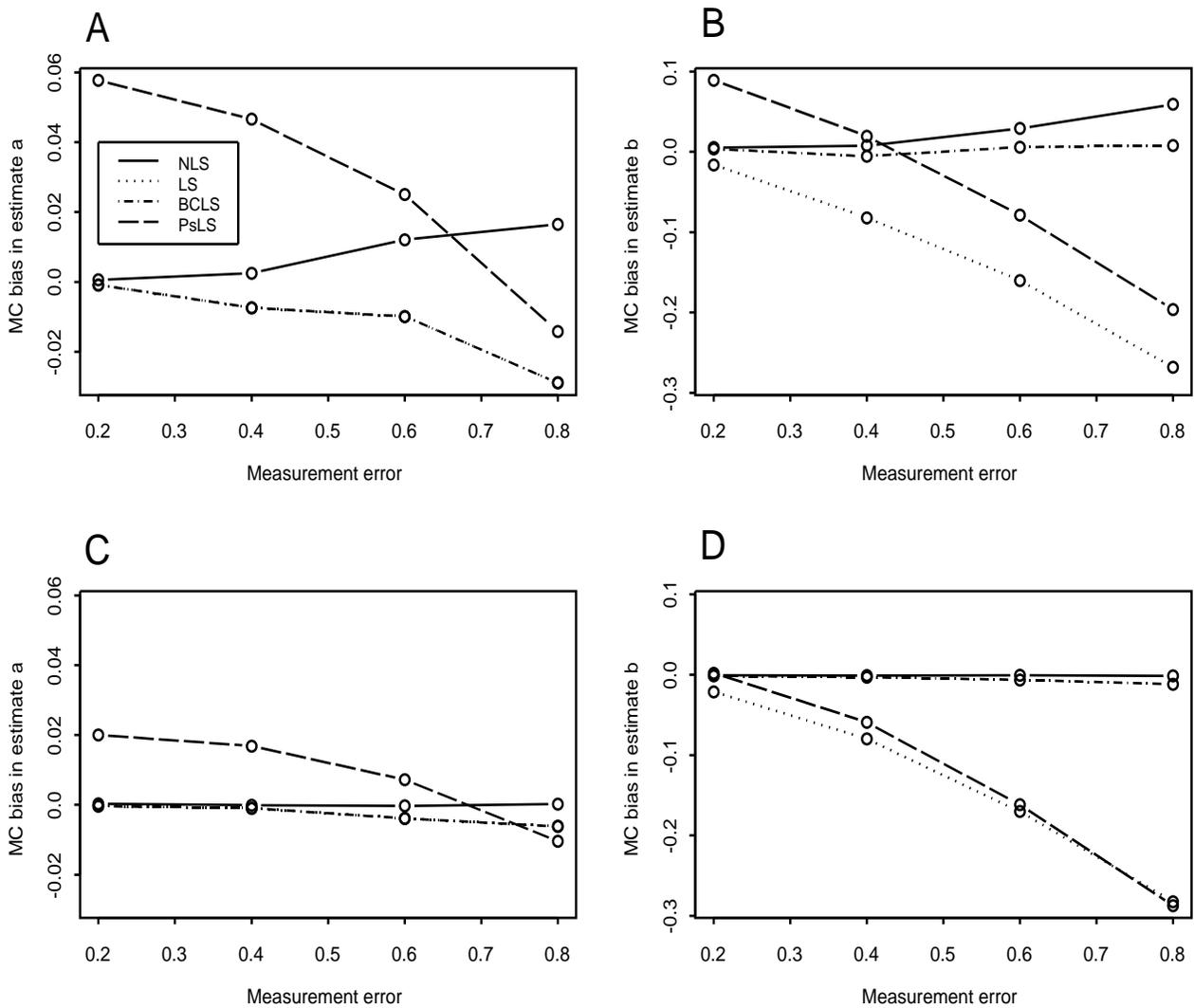,height=6in,width=7in}
}
\end{figure}

\begin{figure} \label{fig3}

\caption{Least squares surfaces for parameters $a$ and $b$ in the Fitzhugh-Nagumo system.
Panels A, B, C: Values $a=0.58$ and $b=0.58$.
Panels C, D, E: Values $a=0.34$ and $b=0.20$}
\centerline{
\epsfig{file=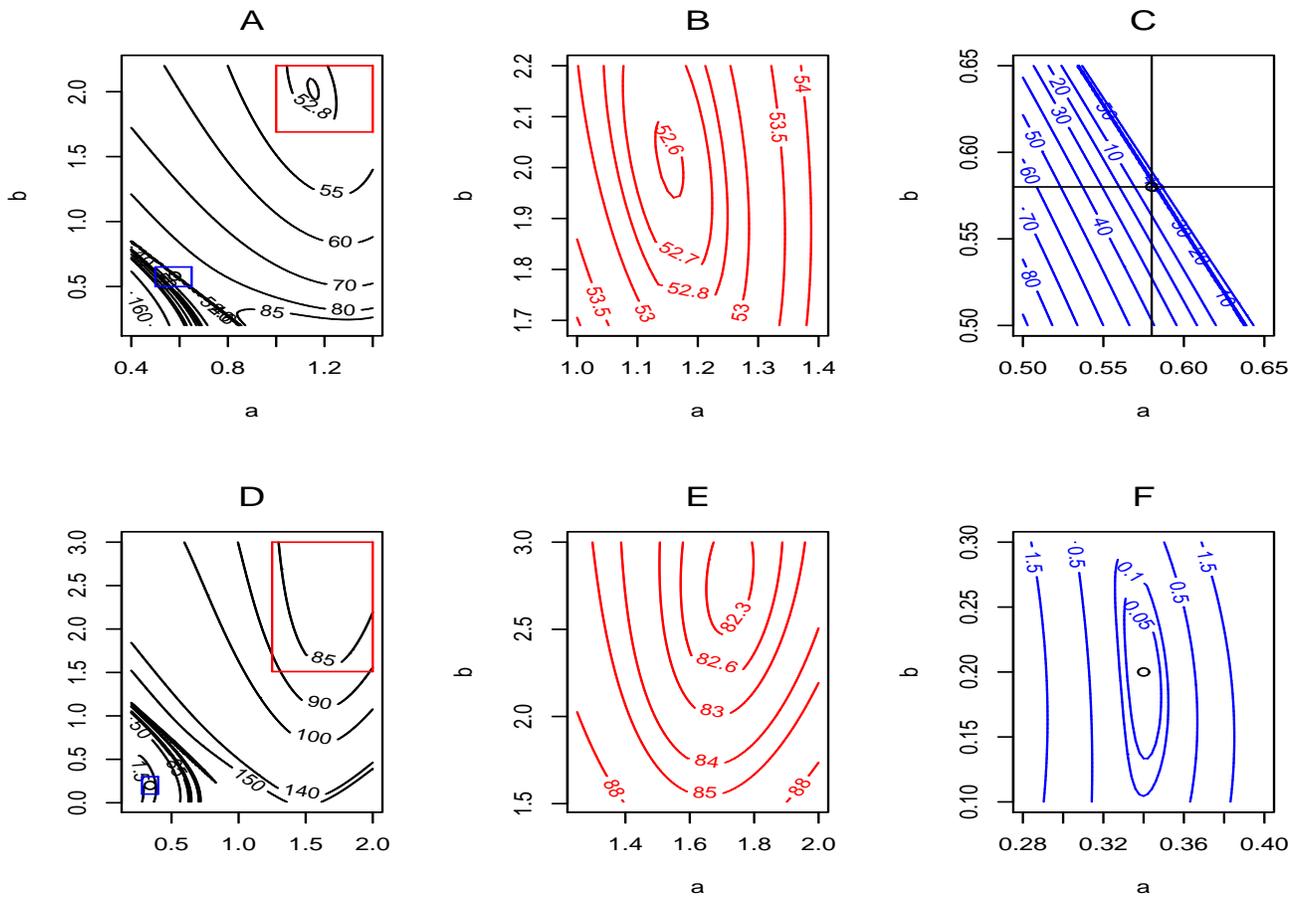,height=5in,width=7in}
}
\end{figure}

\begin{figure} \label{fig4}

\caption{Bias in estimates of $a$, $b$, and $C$ in the Fitzhugh-Nagumo system
when $a=0.34$ and $b=0.2$ (Panels A,B,C) and when $a=0.58$ and $b=0.58$ (Panels C,D,F)
as a function of measurement error.}
\centerline{
\epsfig{file=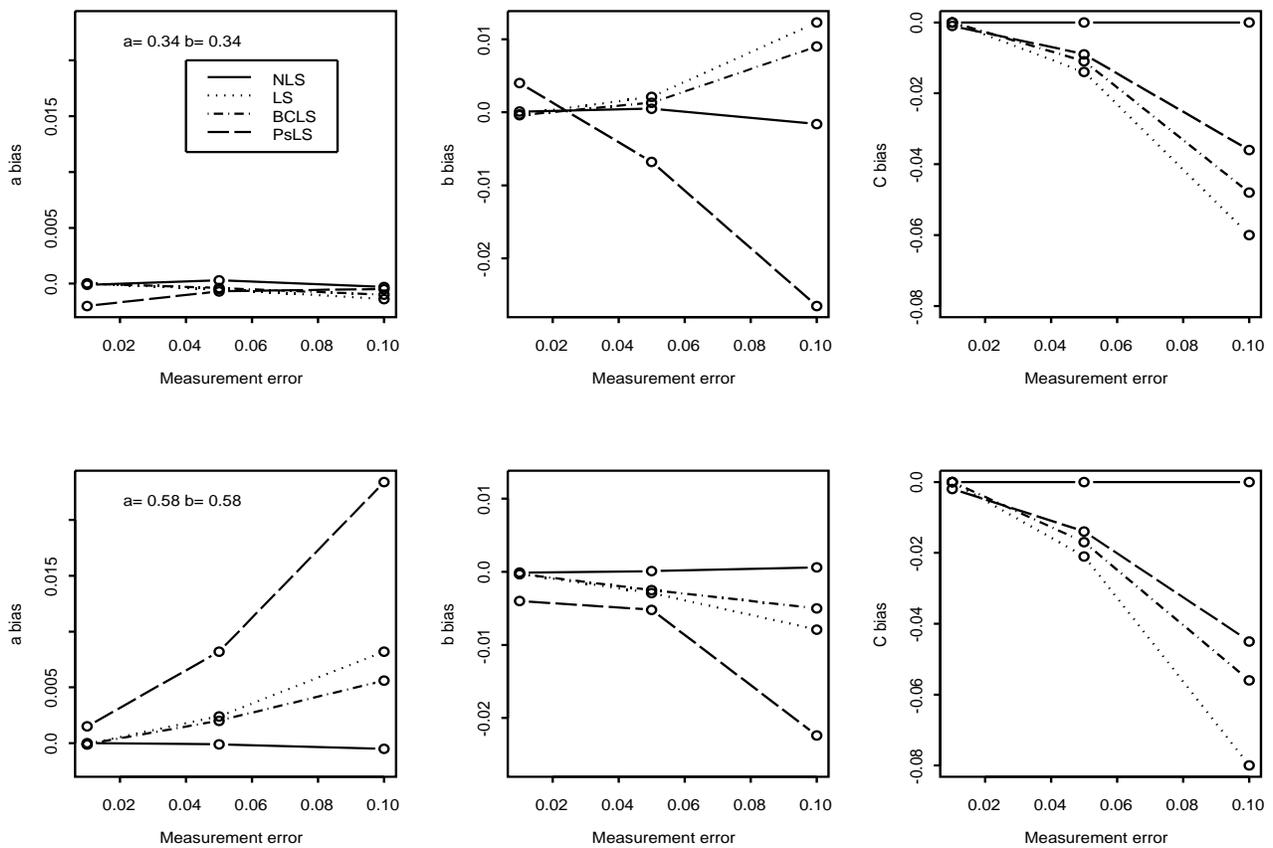,height=5in,width=7in}
}
\end{figure}

\pagebreak

\renewcommand{\arraystretch}{0.92}

\begin{table}
\caption{Results of estimation of parameters in the logistic growth
model to four data sets on growth colonies of the bacteria
Paramecium Aurelium using Nonlinear Least-Square (NLS) and Integrated
Data (BCLS) with bias adjustment methods}
\vspace*{0.25cm}
\begin{center}
\begin{tabular}{l|cc|cc}
\hline
 & \multicolumn{2}{c}{ {\bf NLS} }  &  \multicolumn{2}{c}{ {\bf BCLS} }   \\
\hline
 &  &  & & \\
& {\bf a} & {\bf b } &  {\bf a} &  {\bf b} \\
 &  &  & & \\
\hline
 &  &  & & \\
Data Set  1  &  &  & & \\
Estimate &  0.789  &  0.0014  &  0.776  &  0.0015  \\
 & & & & \\
Parametric &  ( 0.733 , 0.860 )  &  ( 0.0012 , 0.0017 )  &  ( 0.697 , 0.864 )  & ( 0.0012 , 0.0019 )  \\
Bootstrap & & & & \\
Non-Parametric &  &    &  ( 0.738 , 0.839 )  &  ( 0.0014 , 0.0017 )   \\
Bootstrap & & & & \\
 &  &  & & \\
\hline
 &  &  & & \\
Data Set  2   &  &  & & \\
Estimate &  0.837  &  0.0017  &  0.801  &  0.0017  \\
 & & & & \\
Parametric &  ( 0.770 , 0.908 )  &  ( 0.0013 , 0.0021 )  &  ( 0.727 ,0.883 )  & ( 0.0014 , 0.0021 )  \\
Bootstrap & & & & \\
Non-Parametric &  &    &   ( 0.762 , 0.850 )  &  ( 0.0015 , 0.0019 )   \\
Bootstrap & & & & \\
 &  &  & & \\
\hline
 &  &  & & \\
Data Set  3   &  &  & & \\
Estimate &  0.892  &  0.0016  &  0.867  &  0.0017  \\
 & & & & \\
Parametric &  ( 0.810 , 0.989 )  &  ( 0.0012 , 0.0020 )  &  ( 0.746 , 0.971 )  & ( 0.0013 , 0.0024 )  \\
Bootstrap & & & & \\
Non-Parametric &   &    &  ( 0.832 , 0.911 )  &  ( 0.0015 , 0.0019 )   \\
Bootstrap & & & & \\
 &  &  & & \\
\hline
 &  &  & & \\
Data Set  4   &  &  & & \\
Estimate &  0.844  &  0.0016  &  0.817  &  0.0016  \\
 & & & & \\
Parametric &  ( 0.776 , 0.908 )  &  ( 0.0013 , 0.0019 )  &  ( 0.733 , 0.899 )  &
 ( 0.0014 , 0.0018 )
\\ Bootstrap & & & & \\
Non-Parametric &   &    &  ( 0.787 , 0.849 )  &  ( 0.0015 , 0.0017 )  \\
Bootstrap & & & & \\
 &  &  & & \\
\hline
\hline
\end{tabular}
\end{center}
\end{table}

%
\begin{table}
\caption{Simulation Results for Parameters in Logistic Growth Model
using Nonlinear Least-squares (NLS), least squares with (BCLS) and without bias correction and Psuedo-Least square (PsLS) methods with 21 equally spaced observations between $t=0$ and $t=20$. }
\vspace*{0.5cm}
\begin{center}

\begin{tabular}{c l | c c c | c c c c}
\toprule %
 & &  & & & & & &  \\
 &  & \multicolumn{3}{c|}{{\bf a}=0.8} &  \multicolumn{4}{c|}{{\bf b}=0.0015} \\
 & & & & & & & & \\
\hline
& & & & & & & &  \\
  Measurement & & NLS & BCLS & PsLS & NLS &  BCLS  & LS   & PsLS \\
  Error       & &     &    &      &     &     & no bias adj &      \\
& & & & & & & & \\
\hline
  & & & & & & & & \\
0.2 & MC Mean   &  0.801 & 0.799  & 0.846 & 0.00151 & 0.00151 & 0.00148 & 0.00163\\
    & MC s.e     & 0.023 & 0.029  & 0.035 & 0.00011 & 0.00012 & 0.00012 & 0.00018 \\
 & & & & & & & &   \\
\hline
  & & & & & & & &  \\
0.4 & MC Mean   &  0.802 & 0.794  & 0.837 & 0.00151 & 0.00149 & 0.00138 & 0.00153 \\
    & MC s.e.   &  0.046 & 0.057 & 0.072 & 0.00022 & 0.00023 & 0.00022 & 0.00033 \\
 & & & & & & & &  \\
\hline
  & & & & & & & & \\
0.6 & MC Mean   &  0.810 & 0.792  & 0.820 & 0.00154 & 0.00151 & 0.00126 & 0.00138 \\
      & MC s.e  &  0.074 & 0.085  & 0.105 & 0.00035 & 0.00039 & 0.00032 & 0.00045 \\
  & & & & & & & & \\
\hline
 & & & & & & & & \\
0.8 & MC Mean   &  0.813 & 0.777 & 0.789 & 0.00159 & 0.00151 & 0.00110 & 0.00121 \\
    & MC s.e.      &  0.105 & 0.110 & 0.141 & 0.00041 & 0.00056 & 0.00041 &  0.00054 \\
 & & & & & & & &  \\
\hline
\end{tabular}
\end{center}
\end{table}
%

%

\begin{table}
\caption{Simulation Results for Parameters in FitzHugh-Nagamu Model.  Least squares
without bias correction (LS). Bias-corrected least squares (BCLS). Psuedo least squares
with smoothing bandwidth 0.2 (PsLS bw = 0.2). Psuedo least squares with smoothing bandwidth 0.37 (PsLS bw = 0.37).}
\resizebox{\textwidth}{!}{ 
\begin{tabular}{  l  c c   c c   c c   c c   c c  }
\toprule %
Actual  & \multicolumn{2}{c}{{\bf a = 0.34}} & \multicolumn{2}{c}{{\bf  b = 0.2}} & \multicolumn{2}{c}{{\bf C = 3}} & \multicolumn{2}{c}{{\boldmath $v_0$ =-1}} & \multicolumn{2}{c}{{\boldmath $r_0$ =1}}  \\
  &  \multicolumn{9}{c}{  } & \\
\cmidrule(l){2-11}
  & mean &  bias &  mean & bias &  mean  & bias & mean  & bias &  mean  & bias  \\
        & (s.e.) &   &  (s.e.) &   &  (s.e.)  &   & (s.e.)  &   &  (s.e.)  &    \\
\bottomrule %
 LS  &  0.340  &  -0.001  &  0.200  &  0.002  &  2.958  &  -0.014  &  -0.936  &  -0.064  &  1.013  &  0.013  \\
      & ( 0.004 ) &     & ( 0.014 )&         & ( 0.086 ) &          & ( 0.281 ) &       & ( 0.030 ) &  \\
 \cmidrule(l){2-11}
 BCLS &  0.340  &  0.000  &  0.200  &  0.001  &  2.967  &  -0.011  &  -0.984  &  -0.016  &  1.010  &  0.010  \\
      & ( 0.004 ) &      & ( 0.013 ) &       & ( 0.084 ) &           & ( 0.280 ) &      & ( 0.029 ) &  \\
\cmidrule(l){2-11}
 PsLS bw = 0.2 &  0.340  &  -0.001  &  0.199  &  -0.007  &  2.973  &  -0.009  & \multicolumn{2}{c}{NA}    &  \multicolumn{2}{c}{NA}   \\
          & ( 0.024 ) &  & ( 0.042 ) &  & ( 0.045 ) &  &  &  &  &  \\
 \cmidrule(l){2-11}
 PsLS bw = 0.37 &  0.336  &  -0.013  &  0.206  &  0.028  &  3.056  &  0.019  & \multicolumn{2}{c}{NA}    &  \multicolumn{2}{c}{NA}   \\
          & ( 0.013 ) &  & ( 0.029 ) &  & ( 0.046 ) &  &  &  &  &  \\
 \cmidrule(l){2-11}
\toprule 
Actual  & \multicolumn{2}{c}{{\bf a = 0.58}} & \multicolumn{2}{c}{{\bf  b = 0.58}} & \multicolumn{2}{c}{{\bf C = 3}} & \multicolumn{2}{c}{{\boldmath $v_0$ =-1}} & \multicolumn{2}{c}{{\boldmath $r_0$ =1}}  \\
  &  \multicolumn{9}{c}{  } & \\
\cmidrule(l){2-11}
LS  &  0.581  &  0.002  &  0.578  &  -0.003  &  2.938  &  -0.021  &  -0.892  &  -0.108  &  1.020  &  0.020  \\
  & ( 0.005 ) &          & ( 0.014 ) &      & ( 0.118 ) &         & ( 0.248 ) &         & ( 0.041 ) &  \\
 \cmidrule(l){2-11}
BCLS  &  0.581  &  0.002  &  0.579  &  -0.002  &  2.948  &  -0.017  &  -0.966  &  -0.034  &  1.017  &  0.017  \\
     & ( 0.005 ) &        & ( 0.014 ) &     & ( 0.122 ) &          & ( 0.242 ) &         & ( 0.042 ) &  \\
 \cmidrule(l){2-11}
PsLS bw = 0.2 &  0.585  &  0.008  &  0.577  &  -0.005  &  2.959  &  -0.014   & \multicolumn{2}{c}{NA}    &  \multicolumn{2}{c}{NA}   \\
     & ( 0.025 ) &          & ( 0.057) &      & ( 0.060 ) &  &  &  & &  \\
\cmidrule(l){2-11}
PsLS bw = 0.37 &  0.581  &  0.002  &  0.573  &  -0.012  &  3.025  &  0.008   & \multicolumn{2}{c}{NA}    &  \multicolumn{2}{c}{NA}   \\
     & ( 0.016 ) &          & ( 0.036) &      & ( 0.065 ) &  &  &  & &  \\
 \cmidrule(l){2-11}
\bottomrule %
\end{tabular}
  } 
\end{table}

\renewcommand{\arraystretch}{0.95}
\begin{small}
\begin{table}
\caption{Simulation Results for Parameters in FitzHugh-Nagamu Model
using Nonlinear Least-squares (NLS) with various starting values} 
\begin{center}
\begin{tabular}{c c r r r r r r}
\toprule
  \multicolumn{2}{c}{} & {\bf a=0.34} & {\bf b=0.20} &        &  {\bf a =0.58}  &  {\bf b=0.58} & \\
  \cmidrule(l){3-4} \cmidrule(l){6-7}
   \multicolumn{2}{c}{Starting} & \multicolumn{5}{c}{} \\
  \multicolumn{2}{c}{Values} & Mean      & Mean   &  percent    & Mean &  Mean   & percent\\
  a & b &  (s.e.)  &  (s.e.)  & conv.      & (s.e.) &  (s.e.) & conv. \\
\midrule
0.40 & 0.40      &  0.340   & 0.200   & 93.9\%  & 1.154 & 2.018 & 31.7\% \\
     &         &  (0.002) &  (0.015) &       & (0.011) &  (0.032) & \\
     & 0.80     &  0.340   & 0.200    & 94.4\%  & 0.580  & 0.581  & 94.7\% \\
     &        &  (0.002) &  (0.015) &           & (0.007)  &  (0.010) & \\
\cmidrule(l){3-8}
0.80 & 0.40   &  1.720   & 2.771    & 32.6 \%  & 1.155 & 2.018 & 28.5\% \\
     &        &  (0.012) &  (0.043) &         & (0.010)  & (0.031) & \\
     & 0.80     &  1.720   & 2.770    & 33.7\% & 1.154 & 2.017 & 32.1\% \\
     &        &  (0.013) &  (0.043) &     & (0.011) & (0.031)  & \\
\cmidrule(l){3-8}
1.2 & 0.40     &  1.720   &  2.770    & 36.2\% & 1.155 & 2.018 & 31.1\%\\
     &        &  (0.012) &  (0.044) &          &  (0.011) & (0.031) & \\
     & 0.80     &  1.720   & 2.772    & 36.2\%  & 1.155 & 2.016 & 30.2\% \\
     &        &  (0.013) &  (0.043) &           & (0.011)  & (0.030) & \\
\cmidrule(l){3-8}
\multicolumn{2}{c}{From BCLS}    &  0.340  &  0.200  & 94.2\% & 0.580 & 0.581 & 95.1\% \\
     &                         &  (0.002) &  (0.015) &          &  (0.007) & (0.010) & \\
\bottomrule

\end{tabular}
\end{center}
\end{table}
\end{small}


\begin{thebibliography}{50}

\bibitem{Bard:74} Bard, Y. (1974) \newblock \textit{ Nonlinear parameter
    estimation}.
\newblock Academic Press.

\bibitem{Cook:86} Cook R.D, Tsai C.L., Wei B.C. (1986), ``Bias
    in Nonlinear Regression," \textit{Biometrika}, 73, 615--623.

\bibitem{Csa:06} Csajka, C., Verotta, D. (2006)
    ``Pharmacokinetic-Pharmacodynamic Modelling: History and Perspective,"
    \textit{Journal of Pharmacokinetic and Pharmacodynamics}, 33, 227--279.

\bibitem{Cro:86} Crowder, M. (1986) ``On Consistency and Inconcistency of
    Estimating Equations," \textit{Econometric Theory}, 2, 305--330.

\bibitem{Dan:08}  Danhof, M., de Lange, E.C.M, Della Pasqua, O.E,
    Ploeger, B.A, and Voskuyl, R.A. (2008) 11Mechanixm-based
    pharmacokinetic-pharmacodynamic(PK-PD) modeling in translational drug research,"
    \textit{Trends in Pharmacological Sciences}, 29, 186--191.

\bibitem{Dar:07} Dartois, C., Brendel, K., Comets, E., Laffont, C.
    M, Laveliil, C., Tranchand, B., Mentre, F., Lemenuel-Diot, A., and Girard P. (2007)
    ``Overview of model-building strategies in population PK/PD analysis: 2002-2004
    literature survey", \textit{British Journal of Pharmacology},  64, 603--612.

\bibitem{Dig:90} Diggle, PJ. (1990) \textit{Time Series: A Biostatisical
    Introduction}, Oxford University Press:New York.


\bibitem{Efr:93} Efron, B., and Tibshirani, R. J. (1993)
    \textit{ An Introduction to the Boostrap}  Chapman Hall:New York.

\bibitem{Fit:61} FitzHugh, R. (1961) ``Impulses and Physiological States
    in MOndels of Nerve Membrame", \textit{ Biophysical Journal}  1,  445--466.

\bibitem{Foss:71} Foss, S. D. (1971), ``Estimates of chemical kinetic rate
    constants by numerical integration," \textit{ Chemical Engineering Science}, 26,
    485--486.

\bibitem{Fre:80} Freedman, H.I. (1980) \textit{ Deterministic
    Mathematical Models in Population Ecology}  Marcel Dekker:New York.

\bibitem{Gal:04} Galecki A.T, Wolfinger R.D, Linares O.A, Smith
    M.J,Halter J.B. (2004), ``Ordinary Differential Equations PK/PD Models Using the SAS
    Macro NLINMIX," \textit{Journal of Biopharmaceutical Statistics}{ 14}, 483--503.

\bibitem{Gau:34} Gause, GF. (1934) \textit{The Struggle for Existence}
    Willianms and Williams:Baltimore.

\bibitem{HimJon:67}  Himmelblau, D. M.,  Jones, C.
    R., and  Bischoff, K. B. ``Determination of rate constants for complex kinetics
    models," \textit{Industrial and Engineering Chemistry Fundamentals}, 6, 539--543,
    1967.

\bibitem{Ho:95} Ho, D. D., Neumann, A. U., Perelson, A. S., Chen, W.,
    Leonard, J. M., and  Markowitz M. (1995), ``Rapid turnover of plasma virions and CD4
    lymphocytes in HIV-1 infection," \textit{Nature}, 373, 123--126.

\bibitem{Hosten:79} Hosten, L. H. (1979)  ``A comparative study of short
    cut procedures for parameter estimation   in differential equations,"
    \textit{Computers and Chemical Engineering}, 3, 117--126.

\bibitem{Hod:52} Hodgkin, A. L., and Huxley, A. F. (1952) ``A
    Quantitative Description of Membrane Current and Its Application to Conduction and
    Excitation in Nerve," \textit{Jorunal of Physiology} { 133}, 444--479.

\bibitem{Jac:74} Jaquez JA.  (1974) \textit{Compartmental Analysis in
    Biology and Medicine} Elsevier:New York.

\bibitem{Lia:08} Liang, H. and Wu, H. (2008) ``Parameter Estimation
    for differential Equation Models Using a Framework of Measurement Error in Regression
    Models," \textit{Journal of the American Statistical Association}, {103}, 1570--1583.

\bibitem{Lor:63} Lorenz, E.N. (1963) ``Deterministic Nonperiodic Flow"
    \textit{Journal of Atmospheric Sciences}, 26, 130--141.


\bibitem{Mag:03} Mager, D. E, Wyska, E., and Jusko, W. J.
    ''Minireview Diversity of Mechanism-Based Pharmacodynamic Models," (2003), \textit{
    Drug Metabolism and Disposition } { 31}, 510--519.

\bibitem{Nag:62} Nagumo, J. S., Arimoto, S.,
    Yoshizawa, S. (1962), ``An Active Pulse Transmission Line Simulating a Nerve Axon,"
    \textit{ Proceedings of the IRE}{ 50}, 2061--2070.


\bibitem{Per:96} Perelson, A. S., Neumann, A. T., Markowitz, M.,
    Leonard, J. M., Ho, D. D. (1996), ``HIV--1 dynamics in vivo: virion clearance rate,
    infected cell life--span, and viral generation time," \textit{Science}, {  271},
    1582--1586.

\bibitem{Per:97} Perelson, A. S., Essunger, P., Cao, .Y.,
    Vesanen, M., Hurley, A., Saksela, K., Markowitz, M., Ho, D. D. (1997), ``Decay
    characteristics of HIV--1 infected compartments during combination therapy,"
    \textit{Nature}  { 387}, 188--191.



\bibitem{Pre:86} Press, W. H., Teukolsky, S. A., Vetterline, W. T.,
    Flannery, B. P. (1986), \textit{Numerical Recipes in Fortran} Cambridge University
    Press:New York.

\bibitem{Ram:96} Ramsay, J. O. (1996) ``Principal Differential Analysis:
    Data Reduction of Differential Operators," \textit{Journal of the Royal Statistical
    Society, Ser. B} { 58}, 495--508

\bibitem{RamSil:05} Ramsay, J. O. and Silverman, B. W.
    (2005) \textit{Functional Data Analysis} (2nd ed.), Springer:New York.

\bibitem{RHCC:07} Ramsay, J. O., Hooker, G., Campbell, D., and Cao,
    J. (2007), ``Parameter Estimation for Differential Equations: A Generalized Smoothing Approach (with discussion)," \textit{Journal of the Royal Statistical Society, Ser. B}, { 69}, 741--96.

\bibitem{She:00} Sheiner, L.L. and Steimer, J.L. (2000)
    P``harmacokinetic/Pharmacodynaic Modeling in Drug Development," \textit{Annual Reveiw
    of Pharmacological Toxicology}, { 40},  67--95

\bibitem{SwaBre:75} Swartz, J. and Bremermann, H. (1975)
    ``Discussion of parameter estimation in biological modeling: Algorithms   for
    estimation and evaluation of the estimates," \textit{ Journal of Mathematical
    Biology}, 1, 241--257.

\bibitem{Tor:04} Tornoe, C. W., Agerso, H., Johsson, E. N., Madsen,
    H., and Nielsen, H. A. (2004) ``Non-linear mixed-effects
    pharmacokinetic/pharmacodynamic modelling in NLME using differential equations,"
    \textit{Computer Methods and Programs in Biomedicine}, { 76}, 31--40.

\bibitem{Varah:82} Varah, J. M. (1982),``A spline least squares method for
    numerical parameter estimation in   differential equations", \textit{ SIAM, Journal
    of Scientific and Statistical Computation}, 3, 28--46.


\bibitem{Wei:95}  Wei, X., Ghosh, S. K., Taylor, M. E.,  Johnson, V.
    A. Emini, E. A., Deutsch, P., Lifson, J. D., Bonhoeffer, S., Nowak, M. A., Hahn, B.
    H., Saag, M. S., Shaw, G. M. (1995),``Viral dynamics of HIV-1 infection," \textit{
    Nature}  {373}, 117--122.

\bibitem{Wikstrom:97a} Wikstrom, G. (1997) ``Computation of parameters
    occurring linearly in systems of ordinary   differential equations, part i,"
    Technical report, Department of Computing Science, Umea University, 1997.

\bibitem{Wikstrom:97b} Wikstrom, G. (1997) ``Computation of parameters
    occurring linearly in systems of ordinary  differential equations, part ii,"
    Technical report, Department of Computing Science, Umea University,
  1997.

\bibitem{Wilson:99} Wilson, H. (1999) \textit{ Spikes, Decisions and
    Actions: the Dynamical Foundations of
  Neuroscience} Oxford University Press, Oxford, England.


\bibitem{Zha:05} Zhang, W.B. (2005) \textit{Differential Equations,
    Bifurcations, and Chaos in Economics } World Scientific Publishing Company.

\end{thebibliography}
\end{document}